\documentclass[aps,prx,twocolumn,floats,11pt,reprint,floatfix]{revtex4-2}
\usepackage{graphicx,epsfig}
\usepackage{graphics,dcolumn,bm,epic, eepic,fleqn,float}
\usepackage{amssymb,amsmath,amsfonts,multirow,rotate,color}
\usepackage{soul}
\usepackage{wasysym}
\usepackage{color}
\usepackage{xcolor}
\usepackage{wrapfig}
\usepackage{url}
\usepackage{grffile}

\newcommand{\avg}[1]{\langle #1 \rangle}

\newcommand{\bl}{\bullet}

\usepackage{setspace}

\definecolor{darkblue}{rgb}{0,0,0.555}
\definecolor{MyOrange}{rgb}{1,0.77,0.549}
\definecolor{MyPink}{rgb}{1,0.741,0.776}
\definecolor{MyBlue}{rgb}{0.757,0.831,0.902}
\definecolor{MyGreen}{rgb}{0.678,0.859,0.718}
\definecolor{MyLightGreen}{rgb}{0.8,0.859,0.525}

\usepackage[colorlinks=true,
	linkcolor=black,
	urlcolor=blue,
	citecolor=blue,
	bookmarks=true,
	pdfborder={0 0 0}]{hyperref}

\begin{document}
\singlespacing

\title{First-passage times to quantify and compare structural
  correlations and heterogeneity in complex systems}

\author{Aleix Bassolas}
\author{Vincenzo Nicosia}
\affiliation{School of Mathematical Sciences, Queen Mary University of
	London, Mile End Road, E1 4NS, London (UK)}

\begin{abstract}
Virtually all the emergent properties of a complex system are rooted
in the non-homogeneous nature of the behaviours of its elements and of
the interactions among them. However, the fact that heterogeneity and
correlations can appear simultaneously at local, mesoscopic, and
global scales, is a concrete challenge for any systematic approach to
quantify them in systems of different types. We develop here a
scalable and non-parametric framework to characterise the presence of
heterogeneity and correlations in a complex system, based on the
statistics of random walks over the underlying network of interactions
among its units. In particular, we focus on normalised mean first
passage times between meaningful pre-assigned classes of nodes, and we
showcase a variety of their potential applications. We found that the
proposed framework is able to characterise polarisation in voting
systems, including the UK Brexit referendum and the roll-call votes in
the US Congress. Moreover, the distributions of class mean first
passage times can help identifying the key players responsible for the
spread of a disease in a social system, and comparing the spatial
segregation of US cities, revealing the central role of urban mobility
in shaping the incidence of socio-economic inequalities.
\end{abstract}
\date{\today}
\maketitle

The elements of a variety of complex systems can be naturally
associated to one of a small number of classes or categories. Typical
examples include the organisation to which an individual
belongs~\cite{Vanhems2013}, the political party of a
voter~\cite{Fernandez2014}, the income level of a
household~\cite{Jargowsky1996}, or the functional group of a
neuron~\cite{Yan2017}. Quite often, the co-existence of nodes
belonging to different classes and the interactions among those
classes play a fundamental role in the functioning of a system. For
instance, economic and ethnic segregation in cities is known to be
associated to the emergence of social
inequalities~\cite{Barthelemy2016,Batty2017}. Similarly, the
organisation of neural cells in the brain and the intricate patterns
of relations among different functional areas are known to be
responsible for the large variety of cognitive tasks that we as humans
are able to perform~\cite{Bullmore2009,Bassett2017}. However,
obtaining a robust and non-parametric quantification of the
heterogeneity of class distributions, especially in systems consisting
of a large number of interconnected discrete components, is still an
outstanding problem.

We propose here a principled methodology to quantify the presence of
correlations and heterogeneity in the distribution of classes or
categories in a complex system. The method is based on the statistics
of passage times of a uniform random walk on the graph of
interconnections among the units of the system. For instance, in the
case of a social system we can construct a graph among individuals
based on the observed relations or contacts among them. Similarly, in
a urban system we can consider the network of adjacency between census
tracts or the connections among census tracts due to human mobility
flows.  The method we propose moves from the classical research on
mean first passage times between pairs of nodes in a
graph~\cite{Noh2004,Zhang2011,Hwang2012,Bonaventura2014}, and focuses
instead on the distribution of Class Mean First Passage Times (CMFPT)
i.e., the expected number of steps needed to a random walker to visit
for the first time a node of a certain class $\beta$ when it starts
from a node of class $\alpha$. By normalising class mean first passage
times with respect to a null-model where classes are reassigned to
nodes uniformly at random, we can effectively quantify and compare the
heterogeneity of class distributions in systems of different nature,
size, and shape.

We first test the framework on a variety of systems with simple
geometries and ad-hoc class assignments, and then we use it in three
real-world scenarios, namely the quantification of polarisation in the
Brexit referendum and in the US Congress since 1926, the role of
face-to-face interactions among individuals in the spread of an
epidemics, and the relation between economic segregation and
prevalence of crime in the 53 US cities with more than one million
citizens.

\begin{figure*}[!htbp]
  \begin{center}
    \includegraphics[width=6.2in]{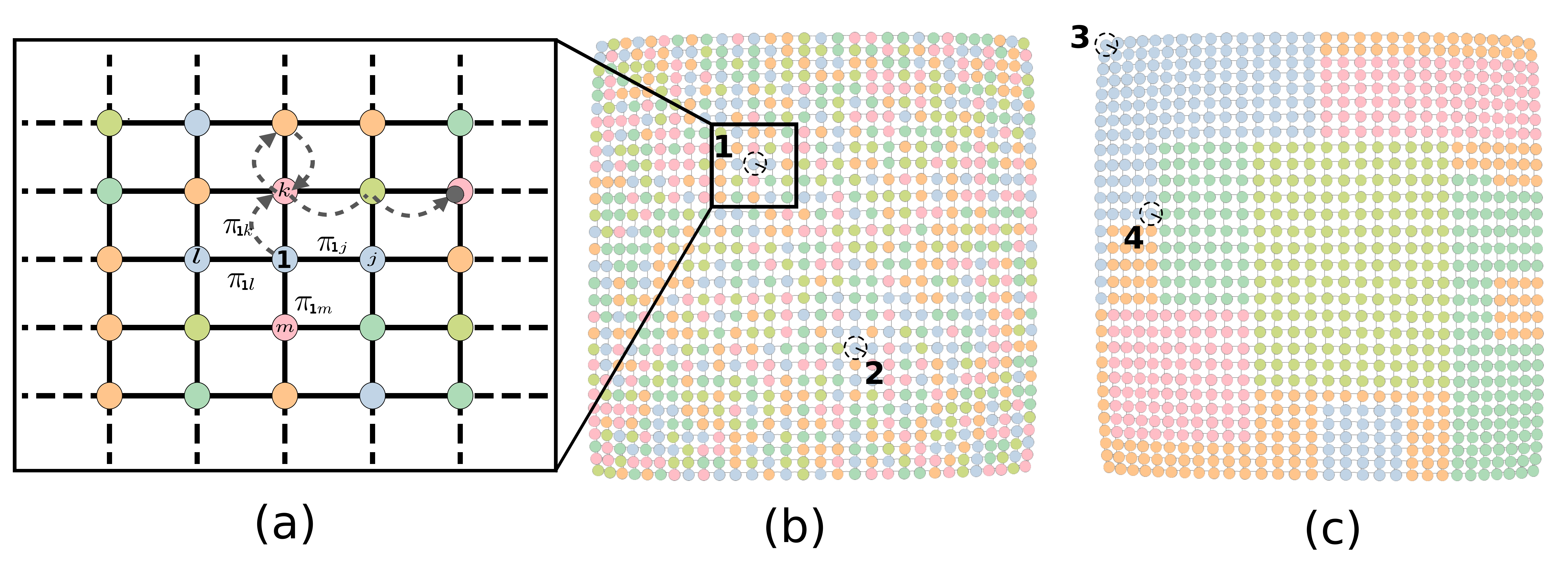}
  \end{center}
  \caption{The statistics of the amount of time needed to a random
    walker to hit a node of a given class when starting from another
    node of the graph encodes useful information about the presence of
    heterogeneity and correlations in the distributions of node
    classes. In panel (a) we show a sketch of the trajectory of a
    random walker over the grid.  In panel (b) the nodes of a square
    lattice are assigned to one of five available classes uniformly at
    random. As a consequence, the time needed to a random walker to
    hit an orange node when starting by node $1$ is comparable to the
    time needed when starting from node $2$. In panel (c) the
    assignment of nodes to classes creates homogeneous clusters of
    nodes of the same colour. As a consequence, the time needed to a
    random walker to hit an orange node when starting from node $3$ is
    much higher than the time needed when starting from node $4$.}
  \label{fig:fig1}
\end{figure*}

\section{Model}

Let us consider a graph $G(V,E)$ with $|V|=K$ edges on $|V|=N$ nodes
with adjacency matrix $A=\{a_{ij}\}$, and a given colouring function
$f:V\rightarrow \mathcal{C}, \quad \mathcal{C} = \{1, 2, \ldots, C\}$,
that associates each node $i$ of $G$ to a discrete label $f(i) =
c_i$. Let us also consider a random walk on $G$, defined by the
transition matrix $\Pi = \{\pi_{ij}\}$ where $\pi_{ij}$ is the
probability that a walker at node $i$ jumps from node $i$ to node $j$
in one step (see Fig.~\ref{fig:fig1}(a)). In general, $\Pi$ could be
any row-stochastic transition matrix, but in the following we will
consider only uniform random walks. We are interested here in the
statistical properties of the symbolic dynamics
$W_i=\{c_{i0},c_{i1},\ldots\}$ of node labels or classes visited by
the walk $W$ when starting from $i$. This dynamics obviously contains
information about the existence of correlations and heterogeneity in
the distribution of colours.

In the regular square lattice shown in Fig.~\ref{fig:fig1}(b), for
example, a function $f$ associates 5 distinct colours to nodes of the
graph uniformly at random, while in Fig.~\ref{fig:fig1}(c), instead,
the colouring function induces a small number of homogeneous
clusters. We expect that, for long-enough times, all the trajectories
of random walks starting from each of the $N$ nodes of the graph in
panel (b) will be associated with the same symbolic dynamics, thus
becoming indistinguishable. Indeed, that system has neither structural
heterogeneity, since all the nodes have the same degree (except for
the nodes at the border of the grid), nor inhomogeneity in class
assignments, since the probability for a node to belong to a certain
class does not depend on its position in the graph or on the classes
of its neighbours. In particular, a random walker starting at any node
belonging to, say, the light-blue class (see nodes $1$ and $2$ in
panel (b)) will require on average a small number of steps to hit a
node in the orange cluster, since orange nodes can be found in the
vicinity of any blue node.

If the colouring function induces compact clusters of nodes belonging
to the same class, as in the lattice shown in Fig.~\ref{fig:fig1}(c),
then the statistical properties of the symbolic dynamics $W_i$ will
heavily depend on the starting node $i$, despite the fact that almost
all the nodes have identical degree. In particular, a random walker
starting in the blue cluster at the top-left corner of the graph (node
$3$) will in general require a very large number of steps before
hitting for the first time an orange node. Conversely, a random walker
starting at node $4$ will hit a node in the orange cluster in a much
smaller number of steps, just because one of its immediate neighbours
is indeed orange.

\subsection{Class Mean First Passage Time}

A quantity of interest in the study of a symbolic dynamics over graphs
is the expected time needed to hit a node of a given colour $\alpha$
for the first time, usually know in the literature as the hitting time
or mean first passage time (MFPT) to class
$\alpha$~\cite{Redner2001}. We denote as $T_{i,\alpha}$ the average
mean first passage time from node $i$ to any node of class $\alpha$,
i.e., the expected number of steps needed to a walk starting on $i$ to
visit for the first time any node $j$ such that $f(j) =
\alpha$. Following the formalism to derive the mean first passage
times between nodes in a graph, we can write a set of self-consistent
equations for $T_{i,\alpha}$~\cite{Masuda2017}:
\begin{equation}
  T_{i,\alpha} = 1 + \sum_{j=1}^{N}
  \left(1-\delta_{f(j),\alpha}\right) \pi_{ij} T_{j,\alpha}.
  \label{eq:T_node_to_class}
\end{equation}
We can derive an analytic solution for Eq.~(\ref{eq:T_node_to_class})
which depends only on the structure of the graph and on the colouring
function $f$.  Let us denote as $T_{\alpha}$ the column vector of
hitting times from nodes of class $\beta\neq\alpha$ to nodes of class
$\alpha$. By convention we set $\{T_{\alpha}\}_i=0$ if
$f(i)=\alpha$. The self-consistent equation for hitting times can be
written as:
\begin{equation*}
  T_{\alpha} = D_{\overline{\alpha}} + \Pi_{\overline{\alpha}}
  T_{\alpha}
\end{equation*}
where $\Pi_{\overline{\alpha}}$ is the transition matrix of the walk
where all the rows and columns corresponding to nodes of class
$\alpha$ are set to zero. We denote by $D_{\alpha}$ the indicator
vector of nodes belonging to class $\alpha$, and by
$D_{\overline{\alpha}} = \bm{1}_N - D_{\alpha}$ the indicator vector
of nodes not belonging to class $\alpha$, i.e.,
$\{D_{\overline{\alpha}}\}_i=1 - \delta_{f(i),\alpha}$. This leads to
the solution:
\begin{equation}
  T_{\alpha} = \left[I - \Pi_{\overline{\alpha}}
    \right]^{-1}D_{\overline{\alpha}}
  \label{eq:hit_time}
\end{equation}
The average Class MFPT $\tau_{\alpha\beta}$ from class $\alpha$ to
class $\beta$ is computed as:
\begin{equation}
  \tau_{\alpha\beta} = \frac{1}{N_{\alpha}}\sum_{j=1}^{N}
  D_{\alpha}^{\intercal}T_{\beta}
  \label{eq:hit_time_avg}
\end{equation}
where $N_{\alpha}$ is the number of nodes belonging to class $\beta$.

The return time to class $\alpha$, that is the expected number of
steps needed to a walker starting on a node of class $\alpha$ to hit a
node of class $\alpha$ (including its starting point) can be computed
in a similar way.  The forward equation for the hitting time to class
$\alpha$ from a node of class $\alpha$ reads:
\begin{equation}
  R_{i,\alpha} = \sum_{j=1}^{N} \delta_{f(j),\alpha} \pi_{ij} +
  \sum_{j=1}^{N} (1-\delta_{f(j),\alpha})\pi_{ij}(1 + T_{j,\alpha})
  \label{eq:R_node_to_class}
\end{equation}
where the first contribution accounts for the neighbours of node $i$
which actually belong to class $\alpha$, while the second contribution
corresponds to walks passing through immediate neighbours of $i$ not
belonging to class $\alpha$. The equation can be written in a compact
form as:
\begin{equation}
  R_{\alpha} = \Pi_{\alpha\alpha}D_{\alpha} + \Pi_{\alpha\overline{\alpha}}
  \left[T_{\alpha}+ D_{\overline{\alpha}}\right]
  \label{eq:return_time}
\end{equation}
where $R_{\alpha}$ is the vector of return times to class $\alpha$,
such that $\left\{R_{\alpha}\right\}_i=0$ if $f(i)\neq \alpha$, and
$T_{\alpha}$ is the vector of MFPT to class $\alpha$ from nodes that
do not belong to class $\alpha$, as above. Here we denote by
$\Pi_{\alpha\alpha}$ is the transition matrix of the walk restricted
to nodes of class $\alpha$, i.e., whose generic element $\pi_{ij}$ is
set to $0$ if either $i$ or $j$ do not belong to class
$\alpha$. Similarly, $\Pi_{\alpha\overline{\alpha}}$ is the transition
matrix restricted to links from nodes of class $\alpha$ to nodes not
in class $\alpha$. By solving Eq.~(\ref{eq:hit_time}) and
Eq.~(\ref{eq:return_time}) for the grid lattice shown in
Fig.~\ref{fig:fig1}(b) we obtain the distribution of CMFPTs:
\begin{center}
  \begin{tabular}{l|ccccc}
    $\tau_{\alpha\beta}$ & $\textcolor{MyBlue}{\bl}$ & $\textcolor{MyLightGreen}{\bl}$ 
    & $\textcolor{MyOrange}{\bl}$ & $\textcolor{MyPink}{\bl}$ 
    & $\textcolor{MyGreen}{\bl}$  \\\hline
    $\textcolor{MyBlue}{\bl}$ & 5.14 & 8.60 & 8.03 & 8.20 & 7.02 \\
    $\textcolor{MyLightGreen}{\bl}$ & 7.91 & 5.58 & 7.82 & 7.96 & 7.18\\
    $\textcolor{MyOrange}{\bl}$ & 7.62 & 8.24 & 4.89 & 8.68 & 7.20 \\
    $\textcolor{MyPink}{\bl}$ & 7.82 & 8.20 & 8.87 & 4.77 & 7.12 \\
    $\textcolor{MyGreen}{\bl}$ & 7.36 & 8.46 & 7.94 & 8.05 & 4.69 \\
    \hline
  \end{tabular}
\end{center}
while for the lattice with clusters in Fig.~\ref{fig:fig1}(c) we get:
\begin{center}
  \begin{tabular}{l|ccccc}
    $\tau_{\alpha\beta}$ & $\textcolor{MyBlue}{\bl}$ & $\textcolor{MyLightGreen}{\bl}$ 
    & $\textcolor{MyOrange}{\bl}$ & $\textcolor{MyPink}{\bl}$ 
    & $\textcolor{MyGreen}{\bl}$  \\\hline
    $\textcolor{MyBlue}{\bl}$ & 5.1 & 255.5 & 183.8 & 263.7 & 158.6 \\
    $\textcolor{MyLightGreen}{\bl}$ & 271.2 & 4.8 & 127.6 & 222.3 & 116.8\\
    $\textcolor{MyOrange}{\bl}$ &  255.4 & 201.5 & 5.0 & 150.6 & 179.3\\
    $\textcolor{MyPink}{\bl}$ &  302.4 & 200.4 & 51.4 & 4.9 & 211.3\\
    $\textcolor{MyGreen}{\bl}$ & 293.1 & 142.1 & 83.5 & 326.2 &  4.9\\
    \hline
  \end{tabular}
\end{center}
Notice that, as expected, the CMFPTs in the lattice with compact
clusters are in general much higher than those observed in the same
lattice with uniformly random class assignments. Moreover, both cases
we have in general $\tau_{\alpha\beta}\neq \tau_{\beta\alpha}$, since
CMFPTs depend primarily on the shape and size of clusters and on the
actual fine-grain arrangement of colours.

\subsection{Mean-Field approximation}

The expressions for class mean first passage time and class return
time provided in Eq.~(\ref{eq:hit_time}) and
Eq.~(\ref{eq:return_time}) are exact, but they have the drawback of
being computationally intensive for graphs with a large number of
nodes. It is possible to construct C-class mean-field approximations
of these expressions, by representing the behaviour of all the nodes
of a class with a single node, and looking at the graph of node
classes. If we denote by $\pi_{\alpha\beta}$ the probability for a
walker to jump in one step from any node of class $\alpha$ to any node
of class $\beta$, e.g. expressed as the total fraction of edges from
nodes of class $\alpha$ to nodes of class $\beta$, the general
Mean-Field equation for CMFPT reads:

\begin{equation*}
  T_{\beta\alpha}^{\rm MF} = D_{\overline{\alpha}} +
  \sum_{\gamma\neq\alpha}\pi_{\beta\gamma}T_{\gamma\alpha}
\end{equation*}
which can be written in compact form as:
\begin{equation*}
  T_{\alpha}^{\rm MF} = D_{\overline{\alpha}} +
  \Pi_{\overline{\alpha}}T_{\alpha}^{\rm MF}
\end{equation*}
where by definition $\{T_{\alpha}^{\rm MF}\}_{\alpha} = 0$ and
$\Pi_{\overline{\alpha}}$ is the transition matrix where the row and
column corresponding to class $\alpha$ are set equal to zero as
above. The solution to this equation is:
\begin{equation*}
  T_{\alpha}^{\rm MF} = \left[I  - \Pi_{\overline{\alpha}}\right]^{-1}
   D_{\overline{\alpha}}
\end{equation*}
Notice that this equation is formally identical to
Eq.~(\ref{eq:hit_time}), with the only difference that it deals with
mean first passage times from all other \textit{classes} to class
$\alpha$, while Eq.~(\ref{eq:hit_time}) provides the mean first
passage times from all the \textit{nodes} in other classes to class
$\alpha$. However, the mean-field approximation only retains
information about the average probability of jumping between two
classes in one time-step, which indeed discards all the (possibly
relevant) structural information contained in the underlying
graph. For instance, if we use the mean-field approximation for the
lattice with random colour assignments shown in
Fig.~\ref{fig:fig1}(b), we get the solutions:
\begin{center}
  \begin{tabular}{l|ccccc}
    $\tau_{\alpha\beta}$ & $\textcolor{MyBlue}{\bl}$ & $\textcolor{MyLightGreen}{\bl}$ 
    & $\textcolor{MyOrange}{\bl}$ & $\textcolor{MyPink}{\bl}$ 
    & $\textcolor{MyGreen}{\bl}$  \\\hline
    $\textcolor{MyBlue}{\bl}$ & 5.05 & 5.09 & 4.98 & 5.10 & 5.08 \\
    $\textcolor{MyLightGreen}{\bl}$ & 5.48 & 5.36 & 5.40 & 5.46 & 5.48 \\
    $\textcolor{MyOrange}{\bl}$ & 4.78 & 4.80 & 4.90 & 4.98 & 4.94\\
    $\textcolor{MyPink}{\bl}$ & 4.92 & 4.89 & 5.00 & 4.94 & 4.90 \\
    $\textcolor{MyGreen}{\bl}$ & 4.74 & 4.74 & 4.80 & 4.74 & 4.80\\
    \hline
  \end{tabular}
\end{center}
where the values of CMFPT between different classes are substantially
different from those computed above using Eq.~(\ref{eq:hit_time}).

\begin{figure*}[!tbp]
  \begin{center}
    \includegraphics[width=6.7in]{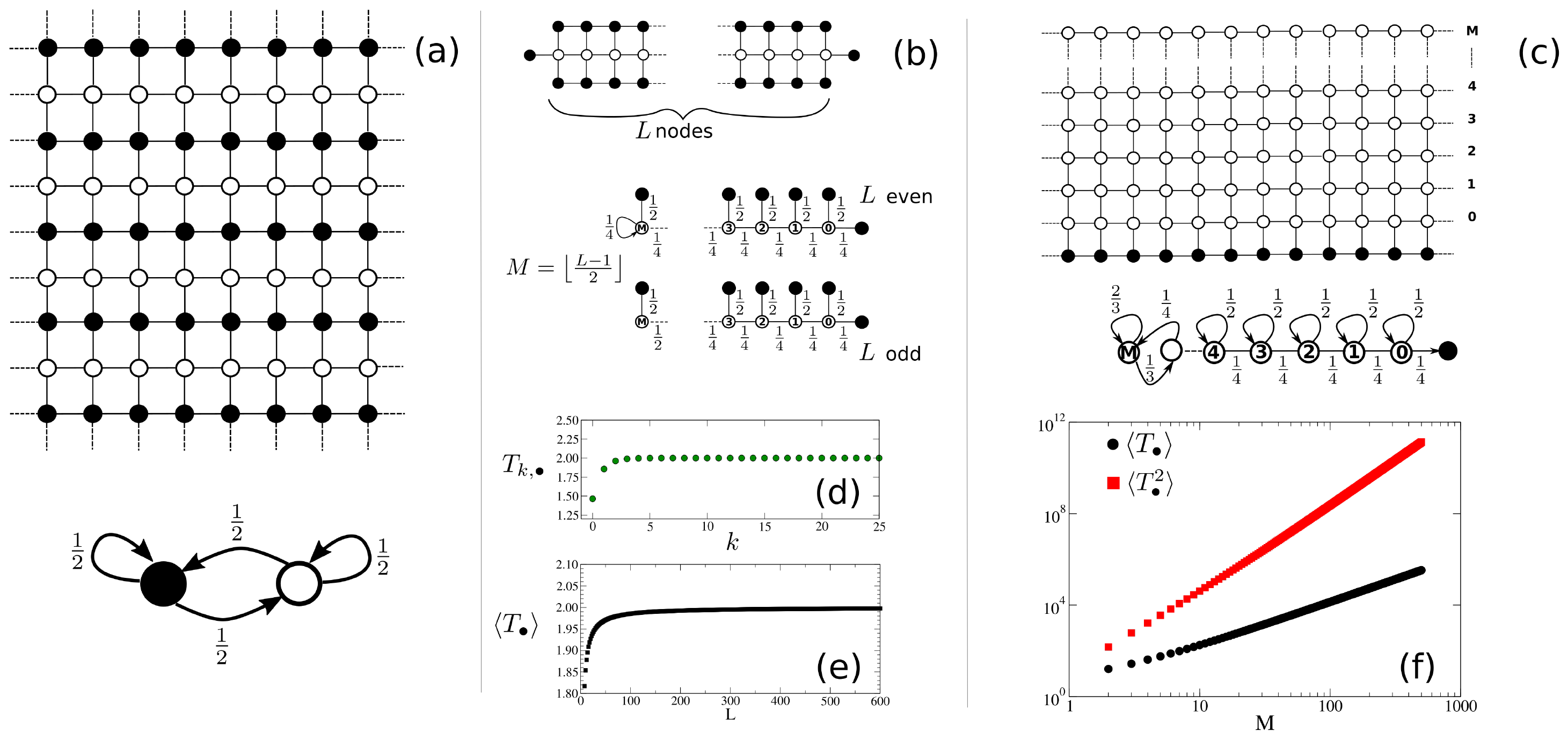}
  \end{center}
  \caption{(a)-(c): Three simple 2-colour geometries on infinite 2D
    lattices (top), and the equivalent minimal graphs used to compute
    inter-class mean first passage times (bottom). When nodes are
    organised in rows of alternate colours (a) the equivalent graph
    contains only two nodes with symmetric connections. The equivalent
    graph of a chain of $\circ$ nodes surrounded by $\bl$ nodes (b) is
    equivalent to a comb graph, while an infinite cylinder of with
    $M+1$ rows of $\circ$ nodes (c) reduces to the chain with $M+1$
    $\circ$ nodes. In the case of the chain graph in (b), the CMFPT of
    a node as a function of its distance $k$ from the endpoint
    converges to $2$ as $k$ increases (d). Similarly, the average
    CMFPT to $\bl$ converges to $2$ when the length $L$ of the chain
    increases (e). In panel (f) we show the first two moments of
    $T_{k,\bl}$ in an infinite cylinder, as a function of the height
    of the cylinder.}
  \label{fig:fig2}
\end{figure*}

\subsection{Normalisation of CMFPT distributions}

As we will see in the following sections, the statistics of CMFPT
depend substantially on the shape, size, and organisation of class
assignments. To allow a fair comparison of CMFPT in systems with
different sizes and shapes, we compute the normalised inter-Class Mean
First Passage Time
$\widetilde{\tau}_{\alpha\beta}=\tau_{\alpha\beta}/\tau^{\rm
  null}_{\alpha\beta}$. This quantity is the ratio between the
expected number of time steps needed to a walk that starts on a node
of class $\alpha$ to reach a node of class $\beta$ for the first time,
and the corresponding value in a null-model where classes are assigned
to nodes uniformly at random, by preserving their relative abundance.
Notice that if classes are distributed uniformly across the system,
then $\widetilde{\tau}_{\alpha\beta}\approx 1, \> \forall
\alpha,\beta$, while in the presence of spatial correlations
$\widetilde{\tau}_{\alpha\beta}$ will deviate from 1.  By using
$\widetilde{\tau}_{\alpha\beta}$ we properly take into account
several common confounding factors, including an uneven abundance of
colours, differences in size and shape, and the effect of borders,
thus making it possible to compare different systems on common
grounds. Depending on the size of the system, the computation of
$\tau_{\alpha\beta}^{\rm null}$ is based on the average over
$10^3-10^5$ independent assignments of classes to the nodes of the
original graph.

\subsection{Computation of CMFPT distributions in real-world networks}

In the following we will compute and use the distribution of CMFPT in
a variety of synthetic and real-world networks. If a system is
naturally represented by an unweighted network, the elements of the
transition matrix of the random walker will be set to
$\pi_{ij}=\frac{a_{ij}}{k_i}$, where $a_{ij}=1$ is there is an edge
from node $i$ to node $j$, and $k_i=\sum_{j} a_{ij}$ is the (out-)
degree of node $i$. If the network is weighted, instead, the elements
of the transition matrix will be $\pi_{ij}=\frac{w_{ij}}{s_i}$, where
$w_{ij}$ is the weight of the edge connecting node $i$ to node $j$,
and $s_i=\sum_{j} w_{ij} $ is the (out-) strength of node $i$. For
networks with up to $N\sim 10^4$ nodes, we solved
Eq.(\ref{eq:hit_time}) and Eq.(~\ref{eq:return_time}) exactly, using
standard linear algebra packages. For larger graphs we reverted
instead to Monte-Carlo simulations, where we estimated the value of
$T_{i,\alpha}$ for each node $i$ of the graph as the average over
$10^4\sim 10^6$ random walks originating at that node, and then
obtained $\tau_{\alpha\beta}$ using Eq.~(\ref{eq:hit_time_avg}).

\section{Simple geometries}

We compute here the distribution of CMFPT for some simple geometries
with simple class assignments. These examples aim at showing that
CMFPT depends heavily on class assignment and on the way nodes
belonging to the same classes are arranged. In all these cases, the
symmetric nature of colour assignments will allow us to perform the
computations on a minimal weighted graph whose distribution of CMFPT
is identical to that of the original system. Notice that those minimal
weighted graphs provide exact solutions for the original colour
assignment they represent, and should not be confused with mean-field
approximations. 

The first example is that of a 2-dimensional square lattice with
periodic boundary conditions (a torus), whose nodes are organised in
alternate stripes of black and white nodes, as shown in
Fig.~\ref{fig:fig2}(a). Thanks to the symmetric nature of this
specific colour assignment, a uniform random walk on that graph is
effectively equivalent to a random walk on the weighted minimal
two-node graph shown in the lower panel of Fig.~\ref{fig:fig2}(a),
with the transition matrix:
\begin{equation*}
  \Pi = 
  \begin{bmatrix}
    \frac{1}{2} & \frac{1}{2}\\
    \frac{1}{2} & \frac{1}{2}\\
  \end{bmatrix}
\end{equation*}
The hitting time to the black node in the minimal equivalent graph
when starting from the white node can be written as:
\begin{equation*}
  T_{\circ,\bl} = 1 + \frac{1}{2} T_{\circ,\bl} 
\end{equation*}
which gives $T_{\circ,\bl} = 2$ and, by symmetry, also $T_{\bl,\circ}
= 2$.

As a second example we consider a finite chain of white nodes
surrounded by black nodes, as shown in the top panel of
Fig~\ref{fig:fig2}(b). We are interested here in showing how the
length $L$ of a linear cluster of a given colour influences the
distribution of CMFPT across the cluster, so we will focus on the
CMFPT from nodes of class $\circ$ to nodes of class $\bl$. In this
case the system has a mirror symmetry, which effectively allows us to
focus on the $\left\lceil\frac{L}{2}\right\rceil$ on either side of
the chain. The only caveat is that weighted minimal graphs associated
to chains with even length $L$ are slightly different from those
associated to chains with odd length, as shown in
Fig.~\ref{fig:fig2}(b). For $L\ge 6$, the closed expression for the
distribution of CMFPT from each node in the chain is:
\begin{equation}
  T_{k,\bl} =
  \left[\frac{4+A_{M-1}}{4-B_{M-1}}\right]\prod_{\ell=1}^{k}B_{M-\ell}
  + \sum_{j=1}^{k-1}A_{M-k+j}\prod_{\ell=0}^{j-1}B_{M-k+\ell}
\end{equation}
where $M=\left\lfloor\frac{L-1}{2}\right\rfloor$, and $A_k$ and $B_k$
are two rational sequences whose form depends on whether the length of
the chain is even or odd. More details about the derivation are
provided in Appendix~\ref{appendix:chain}, where we also compute the
distributions for $1\le L< 6$. In Fig.~\ref{fig:fig2}(d) we report the
distribution of $T_{k,\bl}$ across the chain, while in
Fig.~\ref{fig:fig2}(e) we show the average CMFPT to nodes of class
$\bl$. Notice that when $L\gg 1$, $\avg{T_{\bl}}$ converges towards
$2$, which is the same value obtained in the case of a lattice with
rows of alternate colours seen above, as expected.

As a final example we consider the infinite cylinder shown in the top
panel of Fig.~\ref{fig:fig2}(c), where the nodes in the upper $M+1$
rows are of class $\circ$ and those in the bottom row are of class
$\bl$. The aim of this example is to show the behaviour of CMFPT as a
cluster becomes deeper, i.e., as the nodes in the cluster of a certain
class are placed farther away from the frontier with the other
cluster. For our purposes, this geometry is effectively equivalent to
the linear chain of nodes shown in the bottom half of
Fig~\ref{fig:fig2}(c), where each row of the original graph is
represented by a single node in the minimal graph, with an
appropriately-weighted self-loop. We can write a set of
self-consistent equations for the hitting time to class $\bl$ for a
walk started on each of the rows $k=0,1,\ldots, M$:
\begin{eqnarray}
  T_{0,\bl} = & 2 + \frac{1}{2} T_{1,\bl}\\\nonumber
  T_{k,\bl} = & 2 + \frac{1}{2}\left[T_{k-1,\bl} + T_{k+1, \bl}\right],
  \> k=1,\ldots, M-1\\\nonumber
  T_{M, \bl} = & 3 + T_{M-1, \bl}\\\nonumber
\end{eqnarray}
whose solution is:
\begin{equation}
  T_{k,\bl} = (k+1)(3 + 4M - 2k), \quad k=0, 1, \ldots M-1
 \end{equation}
and
\begin{equation}
  T_{M,\bl} = 3 + M(5 + 2M)
\end{equation}
The full derivation is reported in Appendix~\ref{appendix:cylinder}.
\begin{figure*}[!t]
  \begin{center}
    \includegraphics[width=18cm]{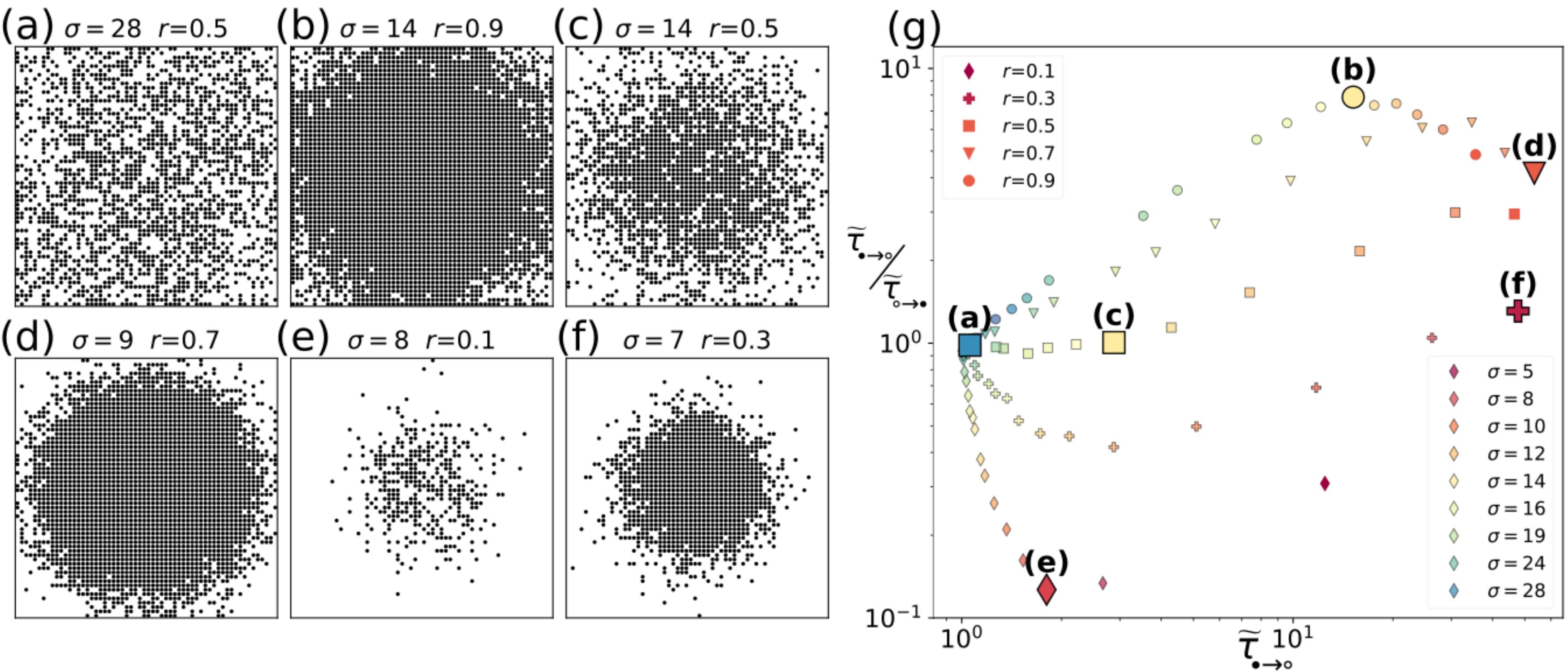}
    \caption{(a)-(f) Snapshots of the model for different values of
      the parameters $\sigma $ and ratio. (a) $\sigma=28$ and
      ratio=0.5 (b) $\sigma=14$ and ratio=0.9 (c) $\sigma=14$ and
      ratio=0.5 (d) $\sigma=9$ and ratio=0.7 (e) $\sigma=8$ and
      ratio=0.1 (f) $\sigma=7$ and ratio=0.3. (g) Transition in the
      coordinates
      $\widetilde{\tau}_{\bullet\to\circ}/\widetilde{\tau}_{\circ\to
        \bullet}$ as a function of $\widetilde{\tau}_{\bullet \to
        \circ}$ for different values of $\sigma$ and the ratio between
      colours. The colour of each point corresponds to the value of
      $\sigma$ in the Gaussian first and type of marker to the ratio
      between colours. Simulations were performed in a regular lattice
      of size $60\times60$.}
    \label{fig:modeltransition}
  \end{center}
\end{figure*}
In Fig.~\ref{fig:fig2}(f) we show the scaling of average mean first
passage time to class $\bl$, defined as:
\begin{align}
  \avg{T_{\bl}} & = \frac{1}{M+1}\sum_{k=0}^{M}T_{k,\bl}= \frac{8M^3 + 33M^2
    + 43M + 18}{6(M+1)}\\
  & \simeq \frac{1}{6}\left(8M^2 + 33M\right)
\end{align}
where the approximation is accurate for $M\gg 1$.  This means that the
average MFPT to class $\bl$ scales as the square of the height of the
cylinder. In the same panel we also show the scaling of the second
moment of $T_{k,\bl}$:
\begin{equation}
  \avg{T_{k,\bl}^2} = \frac{32M^5 + 355M^4}{15(M+1)} + o(M^2) \simeq
  \frac{32M^4 + 355M^3}{15}
\end{equation}
which indicates that the standard deviation of $T_{k,\bl}$ is a
function that grows as $O(M^2)$ as well. In other words, the deeper a
cluster the much higher the CMFPT to its border, and the much wider
the distribution of the MFPT from any single node of the cluster to
the border. 

\section{Synthetic colourings in two-dimensional lattices}

We study here the distribution of CMFPTs in a two-dimensional square
lattice with $N=L\times L$ nodes. Here each node is associated to one
of two possible classes, namely $\bl$ or $\circ$, depending on their
position in the lattice.  Without lack of generality, we set the
relative abundance of $\bl$ nodes $r=\frac{N_{\bullet}}{N}$. Then we
assign $N_{\bullet}$ nodes to class $\bullet$ sampling their
coordinates $(x,y)$ from a symmetric two-dimensional Gaussian
distribution centred in the middle of the lattice, with standard
deviation equal to $\sigma$:
\begin{equation}
P(x) = \frac{1}{{\sigma \sqrt {2\pi } }}e^{{{ - \left( {x - L/2 }
      \right)^2 } \mathord{\left/ {\vphantom {{ - \left( {x - \mu }
            \right)^2 } {2\sigma ^2 }}}
      \right. \kern-\nulldelimiterspace} {2\sigma ^2 }}},
\end{equation} 
By tuning $r$ and $\sigma$ this model allows for a continuous
transition between homogeneous distributions of colours and patterns
with strong class segregation.  In
Fig.~\ref{fig:modeltransition}(a)-(f) we show sample sketches of how
the spatial distribution of colours looks like as a function of
$\sigma$ and $r$. For very large values of $\sigma$, the picture
approaches a homogeneous distribution, regardless of the relative
abundance of the two colours. As $\sigma$ decreases, instead, the
nodes of class $\bl$ will become more strongly clustered around the
centre. The role of the relative abundance of colours is evident from
the comparison of panel (b) and panel (c), which are two
configurations with the same $\sigma$ respectively for $r=0.9$ and
$r=0.5$. In particular, we note that the relative abundance of the two
colours also has a non-trivial role in determining the degree of
mixing between the two classes, since more $\circ$ nodes can be found
inside the $\bl$ cluster for $r=0.5$ than for $r=0.9$.

A summary of the interplay between these two parameters is reported in
Fig.~\ref{fig:modeltransition}(g), where we show the quantities
$\widetilde{\tau}_{\bullet \to \circ}$ and $\widetilde{\tau}_{\bullet
  \to \circ}/\widetilde{\tau}_{\circ \to \bullet}$ for a variety of
values of $r$ and $\sigma$ (the points in panel (g) corresponding to
the configurations in panels (a)-(f) are labelled accordingly). For
large values of $\sigma$, we expect a more homogeneous distributions
of the two colours, (see Fig.\ref{fig:modeltransition}(a)), and indeed
we have $\widetilde{\tau}_{\bullet\to\circ}\simeq 1$, meaning that the
relative distribution of CMFPT from $\bl$ nodes is compatible with the
one observed in the corresponding null model. At the same time, the
ratio
$\widetilde{\tau}_{\bullet\to\circ}/\widetilde{\tau}_{\circ\to\bullet}$
is close to $1$ as well, meaning that the normalised CMFPTs of the two
classes are indistinguishable. As $\sigma$ decreases, the $\bl$
cluster becomes more prominent, but the actual relation between
$\widetilde{\tau}_{\bullet\to\circ}$ and $\widetilde{\tau}_{\bullet
  \to \circ}/\widetilde{\tau}_{\circ\to\bullet}$ will depend on the
value of $r$. In particular, if $r>0.5$, i.e., nodes of class $\bl$
are the majority (see Fig.~\ref{fig:modeltransition}(b) and
Fig.~\ref{fig:modeltransition}(d)), the ratio
$\widetilde{\tau}_{\bullet \to \circ}/\widetilde{\tau}_{\circ \to
  \bullet}$ normally remains larger than $1$. This is again expected,
since a deeper cluster of $\bullet$ nodes causes an increase in the
CMFPT from $\bullet$ to $\circ$ nodes, along the same lines of the
increase in CMFPT observed in the simple geometry shown in
Fig.~\ref{fig:fig2}(c). Conversely, if $r< 0.5$ then $\bullet$ nodes
are interspersed within a large cluster of $\circ$ nodes (see
Fig.~\ref{fig:modeltransition}(e)), making it harder for a walker
started at a $\circ$ node to find a $\bullet$ node. This results in
values of $\widetilde{\tau}_{\circ \to \bullet}$ larger than $1$,
i.e., much longer than in the corresponding null-model, and in
$\widetilde{\tau}_{\bullet\to\circ}/\widetilde{\tau}_{\circ\to\bullet}<1$. There
are some particular situations (see
Fig.~\ref{fig:modeltransition}(c),(f)) in which despite the increase
in $\widetilde{\tau}_{\bullet\to\circ}$, the ratio
$\widetilde{\tau}_{\bullet\to\circ}/\widetilde{\tau}_{\circ\to\bullet}$
remains close to $1$. This is due to the fact that in these cases the
two classes are distributed in a similar fashion, and the relative
depths of the two clusters are indeed comparable.

It is worth noting that the $(\widetilde{\tau}_{\bullet\to\circ},
\widetilde{\tau}_{\bullet\to\circ}/\widetilde{\tau}_{\circ\to\bullet})$
phase space provides a very intuitive interpretation and a powerful
visualisation of the heterogeneity of distributions of two classes,
and it would thus be quite useful to characterise the spatial
heterogeneity of generic binary feature distributions and point
statistics~\cite{Marcon2003,Marcon2010}.

\section{Polarisation and segregation in voting dynamics}

Among the variety of social dynamics that exhibit complex behaviours,
voting is possibly one of the most interesting. And not just because
anything concerning politics can spur endless and ferocious
discussions, but also because voting patterns are the result of the
interplay of a variety of factors that are generally difficult to
model in an accurate way, including socio-economic and cultural
background and spatial and temporal
correlations~\cite{Fernandez2014,Braha2017}. We focus here on two examples where
voting dynamics result in the emergence of heterogeneity and
correlations, namely the spatial clustering of Leave/Remain voters in
the Brexit referendum and the polarisation of opinions of roll-call
votes in the US Congress.

\subsection{Spatial heterogeneity of Brexit vote}

Here we show how inter-class mean first-passage times can be used to
quantify the spatial segregation of voting patterns. We consider the
results of the so-called ``Brexit'' referendum, held in the United
Kingdom in 2016 to decide whether to leave the European Union. The
referendum had a turnout of $72.2\%$, and $51.9\%$ of the voters
expressed the preference to leave the EU. The results of the
referendum have been analysed in several works~\cite{Stolz2016} which
have outlined interesting correlations of voting preference with a
variety of socio-economic indicators, including age, income level,
unemployment, and level of education~\cite{Dlotko2019}. One of the
most intriguing aspects of the results was that the vote was highly
segregated. Indeed, highly urbanised areas, as well as the majority of
constituencies in Scotland, voted preferentially for Remain, while the
rest of the country expressed a preference to Leave.

We constructed the planar graph of constituencies in Great Britain
(i.e., all the mainland constituencies in England, Wales, and
Scotland, leaving out the few constituencies in Northern Ireland),
where each node is associated to a constituency and a link between two
nodes exists if the corresponding constituencies border each other. We
assigned each node to either Leave or Remain according to which party
won the majority of votes in the corresponding constituency. We then
computed the normalised average CMFPT pattern for Remain (R) and Leave
(L), we obtained the values shown in
Table~\ref{table:Brexit}(a).

\begin{table}
  \begin{center}
    \begin{tabular}{cc}
      \begin{tabular}{r|cc}
        $\widetilde{\tau}_{\alpha\beta}$ & L & R\\\hline
        L & 0.976 & 2.857 \\
        R & 12.304 & 1.008 \\
        \hline
      \end{tabular} \hspace{1cm}&\hspace{1cm}
      \begin{tabular}{r|cc}
        $\widetilde{\tau}_{\alpha\beta}$ & L & R\\\hline
        L & 1.018 & 1.057\\
        R & 1.090 & 0.981\\
        \hline
      \end{tabular}\\
      & \\
      (a) \hspace{1cm}&\hspace{1cm} (b) \\
      \end{tabular}
  \end{center}
  \caption{(a) Normalised CMFPT between constituencies which voted for
    Leave (L) or Remain (R) in the Brexit referendum. Note that
    $\widetilde{\tau}_{LR}$ is almost six times smaller than
    $\widetilde{\tau}_{RL}$. (b) If we consider the ensemble of colour
    assignments where a node is assigned to Leave (Remain) with a
    probability equal to the fraction of voters supporting Leave
    (Remain) in that constituency, the resulting normalised CMFPT is
    not distinguishable from the null-model distribution.}
  \label{table:Brexit}
\end{table}
It is worth noting that while the normalised return time to each class
is close to $1$ in both cases, i.e., it is consistent with the
corresponding null model where classes are reassigned to nodes
uniformly at random, the proper inter-class MFPTs exhibit a pronounced
disparity between the two classes. In particular, the normalised CMFPT
from Leave to Remain $\widetilde{\tau}_{LR}$ is much smaller (2.827)
than its counterpart $\widetilde{\tau}_{RL}$ (12.304). This means
that, on average, in this graph is much easier for a random walker
starting at a node whose citizens expressed a majority of votes for
Leave to arrive at a node where people preferentially voted for
Remain, than the other way around. Or, putting it in another way, it
was much easier for a Leave supporter wandering through the graph to
meet a Remain supporter than the other way around.

\begin{figure*}[!t]
  \begin{center}
    \includegraphics[width=18cm]{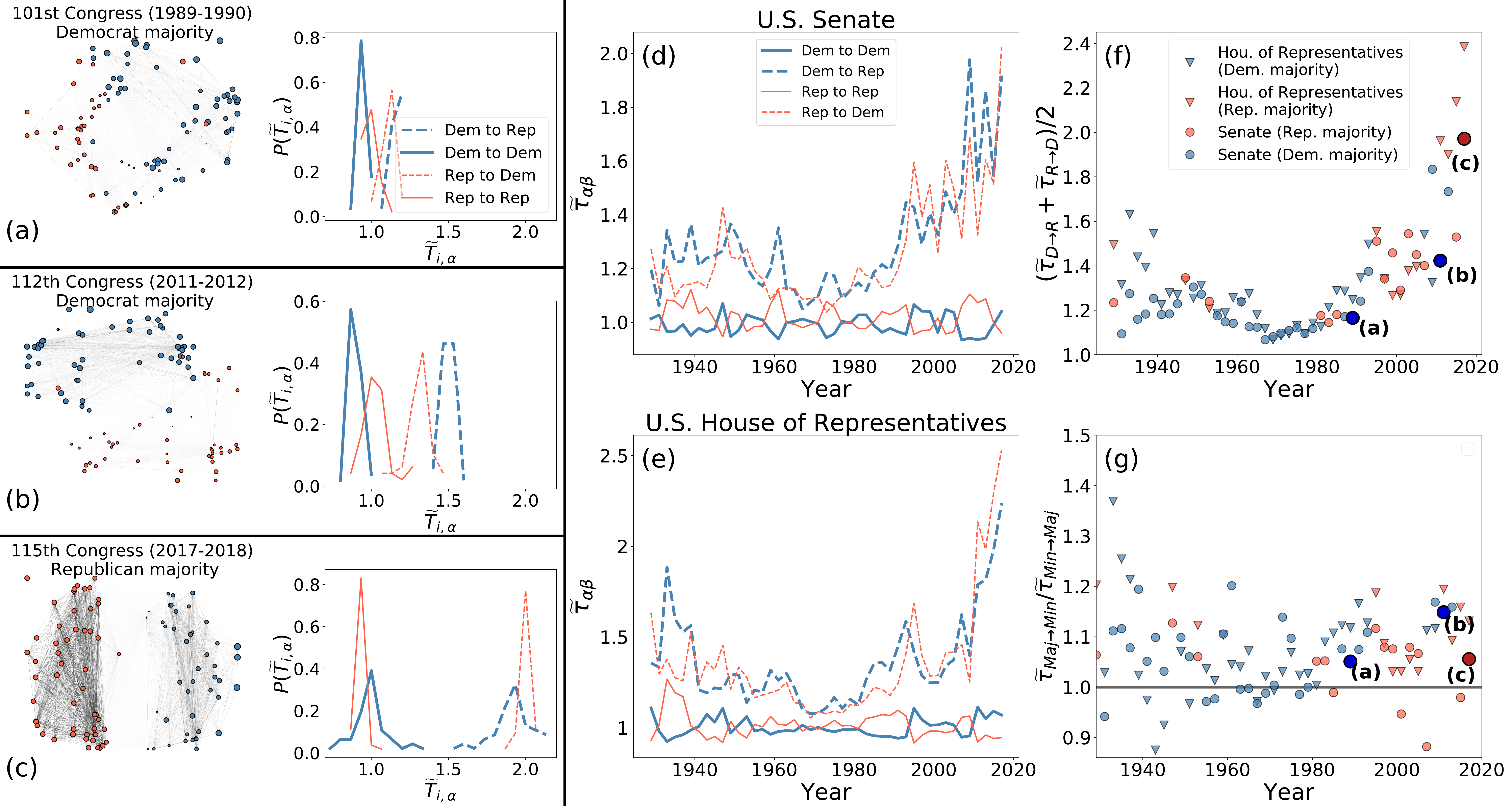}
    \caption{Roll-call polarisation in the Senate and the U.S. House of
      Representatives. (a)--(c) Three networks connecting the members
      of the senate where edge weights correspond to the number of
      calls in which the corresponding two members voted the same on a
      roll call, and the corresponding distribution of CMFPT from the
      nodes (members) to each of the parties. (a) 101st Congress
      (1989-1990), (b) 112th Congress (2011-2012) and (c) 115th Congress
      (2017-2018). (d)-(e) Temporal evolution of $\tau_{\alpha \beta}$
      in (d) the Senate and (e) the U.S. House of Representatives. (f)
      Temporal evolution of polarisation in the the Senate and the
      U.S. House of Representatives, as measured by
      $(\widetilde{\tau}_{D \to R}+\widetilde{\tau}_{R \to D})/2$. (f)
      Temporal evolution of $\widetilde{\tau}_{Maj \to
        Min}/\widetilde{\tau}_{Min \to Maj}$. The fact that this
      quantity is larger than one in most of the terms indicates that
      the party with a majority is the one responsible for the largest
      contribution to polarisation in the corresponding term.}
    \label{fig:rollcalls}
  \end{center}
\end{figure*}

A possible interpretation of this result is the presence of a
structural reinforcement of segregation. Indeed, if we assume that
voters could influence each other by discussing the matter of the
referendum with other voters holding opposite opinions, then a person
determined to vote for Remain would have had a much harder time
finding a Leave supporter to convince. On the contrary, Leave
supporters would have been able to find Remain supporters much more
easily, as Leave constituencies are on average closer (in terms of
MFPT) to other Remain constituencies than the other way around.

However, this picture is completely reversed if we take into account
the actual percentage of votes for Leave and Remain in each
constituency, instead of noting only which party got a majority. We
considered an ensemble of colour assignments obtained by assigning
colour $L$ to node $i$ with probability $p_L(i)$ equal to the
percentage of Leave voters in constituency $i$. For instance, if in a
given constituency $i$ we had $45\%$ of votes for Leave, then node $i$
will be assigned to class $L$ with probability $0.45$, and to class
$R$ with probability $0.55$. We computed the inter-class Mean First
Passage Times in this ensemble of colourings, and normalised them by
the corresponding values in a null-model where we re-assigned vote
proportions among nodes uniformly at random. The results are reported
in Table~\ref{table:Brexit}(b).

It is evident that, by taking into account the actual distribution of
voters in each constituency, the pattern of inter-class MFPT becomes
practically indistinguishable from the one we would observe in the
corresponding null-model. This means that there was indeed no
significative spatial segregation effect in the Brexit vote, and
indicates that indeed the reasons in support for Leave or Remain were
most probably linked to socio-economic characteristics, rather than to
geographical ones.

\subsection{Polarisation of roll-call votes}   
        
In this section we show how $\widetilde{\tau}_{\alpha\beta}$ can be
used to quantify the level of polarisation in roll-call votes in the US
congress~\cite{Waugh2009,Hirano2010,Neal2020}, and to keep track of
its evolution over time. We considered the full data set of affiliation
and single roll-call votes of the members of the US Congress, and we
built a weighted graph among members of each chamber in each term
between 1929--2016, where the weight of the edge connecting two nodes
is equal to the number of times the votes of the corresponding members
in a roll-call coincided. We assigned each node to either Republicans,
Democrats, or Others, according to the party to which they belonged,
but we focused exclusively on the CMFPT between Republicans and
Democrats, since members of Other parties are normally a rather small
minority, if present at all.

It is worth noting that, at difference with the synthetic networks we
have studied so far, these networks do not admit a natural embedding
in a metric space. Intuitively, we expect that members of both the
Senate and the House of Representatives would, in general, be more
likely to vote as other members of their party, giving rise to somehow
definite clusters. However, the situation is not always that clear.
In Fig.~\ref{fig:rollcalls}(a)-(c) we show the networks of Senate
members observed in three different terms from the last 30
years. Indeed, it is evident that stronger connections between members
of the same party appear as time passes. Moreover, when comparing the
112th and 115th congresses, we can see that the party with majority
tends to be more heavily connected than the other one. This is also
reflected in the relative size of nodes, which is proportional to the
corresponding CMFPT to the other party. Just by looking at these three
graphs we would be inclined to think that the level of polarisation in
the US Congress seems to increase over time.

For each graph, we also show the distribution of normalised mean first
passage times $\widetilde{T}_{i, \alpha}$ from each node $i$ to each
of the parties, respectively for Democrats and Republicans.  By
looking at these distributions on the same scale, it is evident that
the polarisation, intended as the relative distance between nodes of
different parties as measured by inter-class MFPT, has increased
substantially in the last thirty years. In particular, the separation
between the intra- and inter-class MFPT distributions has increased
dramatically, to the point that in the 115th Congress the
distributions of intra-class and inter-class MFPT are clearly
separated. When comparing the 112th and 115th legislatures we also
observe that the changes in the shape and position of the MFPT
distributions seem to depend on which party holds the majority of the
seats. In particular, when Democrats have the majority, the CMFPT to
Republicans across nodes is higher than from Republicans to
Democrats. Conversely, when Republicans are ruling the situations is
inverted, pointing out that the party that holds the majority seems to
be the main driver of polarisation.

\begin{figure*}[!ht]
    \begin{center}
    \includegraphics[width=18cm]{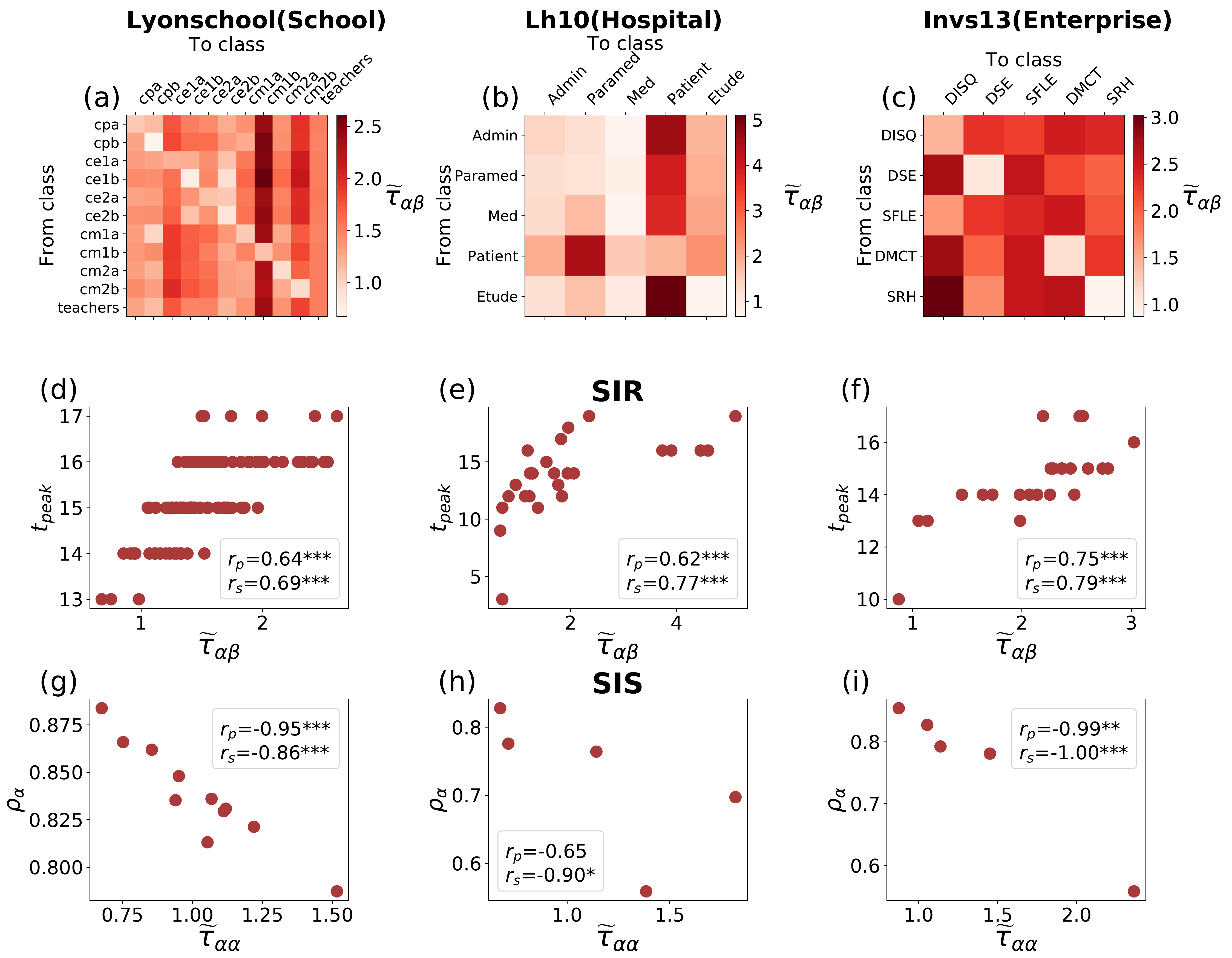}
      \caption{Class mean first passage times and the spread of
        epidemics in contact networks.  (a)-(c) Class mean first
        passage times in contact networks corresponding to a school, a
        hospital and an enterprise.  (d)-(f) Connection between
        $\widetilde{\tau}_{\alpha \beta}$ and the time steps until the
        peak of the epidemic $t_{peak}$ in a SIR model.  (g)-(i) Connection
        between the return times $\widetilde{\tau}_{\alpha \alpha}$
        and the fraction of infected individuals in each class in the
        stationary state. Simulations were performed using the
        parameters $\beta=0.7$ and $\mu=0.1$.} \label{FigSIR}
    \end{center}
\end{figure*}

In Fig.~\ref{fig:rollcalls}(d)-(e), we show the evolution of the
average $\widetilde{\tau}_{\alpha \beta}$ between the two main US
parties in both the Senate and the House of Representatives. The
dashed lines indicate the intra-class MFPT, and indeed support the
intuition that polarisation has increased dramatically in recent
years. The plots of return times $\widetilde{\tau}_{\alpha \alpha}$
instead (solid lines), are far more stable, and the small oscillations
we observe depend only on which is the ruling party, since that one
will normally yield lower values of $\widetilde{\tau}_{\alpha
  \alpha}$.

We quantify the overall polarisation in the Senate and the US House of
Representatives by computing the average between the inter-class
passage times $(\widetilde{\tau}_{D \to R}+\widetilde{\tau}_{R \to
  D})/2$, which provides an estimate of how close are the voting
behaviours of the two main parties in each Congress. The temporal
evolution of $(\widetilde{\tau}_{D \to R}+\widetilde{\tau}_{R \to
  D})/2$ is shown in Fig.~\ref{fig:rollcalls}(f), where each point is
coloured according to the party holding the majority of seats in that
term. Again, we observe a clear increase of polarisation after the
1970s in both branches of the Congress, which is in very good
agreement with prior
works~\cite{Waugh2009,Hirano2010,Neal2020}. Notice that there are some
interesting patterns there, like for instance the term starting in
January 2001, which displays the lowest polarisation in the last 25
years, most likely due to the 9/11 terrorist attacks later that year.

We inspect next if, as hypothesised, the increasing polarisation
observed in Fig.~\ref{fig:rollcalls}(f) is due to the party holding
the majority of seats in each term. We computed the ratio of
inter-class passage times $\widetilde{\tau}_{Maj \to
  Min}/\widetilde{\tau}_{Min \to Maj}$, where $\widetilde{\tau}_{Maj
  \to Min}$ is the normalised inter-class MFPT from the party with the
majority to the party with the minority and $\widetilde{\tau}_{Min \to
  Maj}$ is the normalised inter-class MFPT from the minority to the
majority. The values of $\widetilde{\tau}_{Maj \to
  Min}/\widetilde{\tau}_{Min \to Maj}$ (Fig.~\ref{fig:rollcalls}(g))
are indeed relatively stable over time, with the vast majority of
points lying along or above the solid grid line corresponding to
absence of polarisation.  This indicates that in most of the terms
over the last 80 years the party holding the majority has been the
main driver of polarisation. These results demonstrate that
$\widetilde{\tau}_{\alpha \beta}$ is indeed a robust measure of
polarisation. We argue that the same framework could be easily used in
other contexts, including the polarisation of discussions in (online)
social networks~\cite{Guerra2013,Matakos2017} or the flip of
candidates between different political parties~\cite{Faustino2019}. In
particular, it could be also possible to identify those agents or
individuals who contribute more to polarisation by looking at the
ranking of nodes by their values of CMFPT to other classes.

\section{Contact assortativity and relation with epidemics spreading}

Understanding the mixing between groups of individuals in a network
can provide a lot of information about the properties of social
dynamics, including the role played by different individuals in the
transmission of
diseases~\cite{Starnini2013,Vanhems2013,Barrat2014,Kiti2016}.  As a
second case study, we use CMFPTs to identify how different groups of
people interact in three face-to-face contact networks, namely the
contacts in a hospital, in a school, and in an enterprise, obtained
from the SocioPatterns project data
set~\cite{Sociopatterns,Vanhems2013,Genois2015,Genois2018}.

The definition of groups or classes is specific of each system, e.g.,
the role of a person in the case of hospitals, the class in the case
of schools, and the department in which a person works for the network
of contacts in an enterprise environment. The weight of the undirected
edge connecting two nodes in each graph is equal to the number of
contacts between the corresponding individuals. By looking at the
dynamics of two simple epidemic models, namely a
Susceptible-Infected-Susceptible (SIS) and a
Susceptible-Infected-Recovered (SIR), we show here that the
distribution of CMFPT in each graph provides relevant information
about the dynamics of disease spreading in the system. For both
epidemic models, if an individual $i$ is infected it selects one of
its neighbours $j$ with probability $w_{ij}/\sum_{j}w_{ij}$ and
infects it with a probability $\beta$. Afterwards, each of the
infected individuals will either recover with probability $\mu$ in the
case of the SIR model, or become susceptible again in the case of the
SIS model.

In Fig.~\ref{FigSIR}(a)-(c) we report the matrices of
$\widetilde{\tau}_{\alpha \beta}$ respectively for the school, the
enterprise, and the hospital. In the school network, we observe a
consistent pattern of lower values of normalised intra-CMFPT
$\widetilde{\tau}_{\alpha \alpha}$, which is most probably due to the
much higher number of contacts among individuals in the same classroom
compared to individuals in other classrooms. Notwithstanding this
general pattern, some of the classes are more tightly connected than
other close-by classes, as in the case of ce1b and ce2b. Also in the
case of the enterprise network we distinguish a clear pattern where
for each class $\alpha$ the value of $\widetilde{\tau}_{\alpha\alpha}$
is normally much smaller than $\widetilde{\tau}_{\alpha\beta}$ for
$\alpha\neq\beta$. Yet, there are some interesting deviations, as in
the case of SFLE. This department plays a similar role to
that of teachers in the case of the school network (i.e., contain
people that tend to interact with more than one class), and all the
other departments exhibit similar values of inter-CMFPT to that
class. Finally, in the hospital contact network patients seem to 
be the most isolated, while the paramedical staff and
display lower CMFPT to all the other classes.

As we show in the following, we found a quite interesting relation
between the pattern of CMFPT in each network and the spreading
dynamics on the same graph. We run a large number of simulations of
the SIR and SIS models, seeding the disease in each of the nodes of a
network, and calculating the number of time-steps needed to the spread
to reach the peak in each of the classes, as a function of the group
to which the seed node belongs. In Fig.~\ref{FigSIR}(d)-(f) we report,
as a function of $\widetilde{\tau}_{\alpha \beta}$, the average number
of steps until the peak of the epidemic $t_{peak}$ in class $\beta$ is
reached in a SIR model where the seed is a node in class
$\alpha$. Interestingly, we found that the time to the peak in class
$\beta$ is an increasing function of $\widetilde{\tau}_{\alpha
  \beta}$, and the rank correlation between the two variables is
always pretty large and significant in the three systems.

The results on the SIS model somehow complement the picture observed
in the case of SIR. In Fig.~\ref{FigSIR}(g)-(i) we show that the class
mean return times $\widetilde{\tau}_{\alpha \alpha}$ are strongly
correlated with the fraction $\rho_{\alpha}$ of infected individuals
of that class in the stationary state. In particular, the lower the
value of $\widetilde{\tau}_{\alpha \alpha}$, the larger the fraction
of infected individuals in class $\alpha$ in the endemic state,
suggesting that the steady-state dynamics is indeed predominantly
driven by interactions among individuals in the same class. In
Table~\ref{Table2} we report these correlations for a wider range of
$\beta$ and $\mu$. As shown in detail in Fig.~\ref{FigSIRtot} and
Fig.~\ref{FigSIRnorm}, the observed correlations between
$\rho_{\alpha}$ and $\tau_{\alpha\beta}$ are consistently higher than
the correlation with either the total number of edges between two
classes or the total fraction of edges from class $\alpha$ to class
$\beta$. Those results as well as the correlations with a wider range
of parameters and contact networks can be found in
Appendix~\ref{appendix:epidemics}.

\begin{figure*}[!th]
    \begin{center}
    \includegraphics[width=18cm]{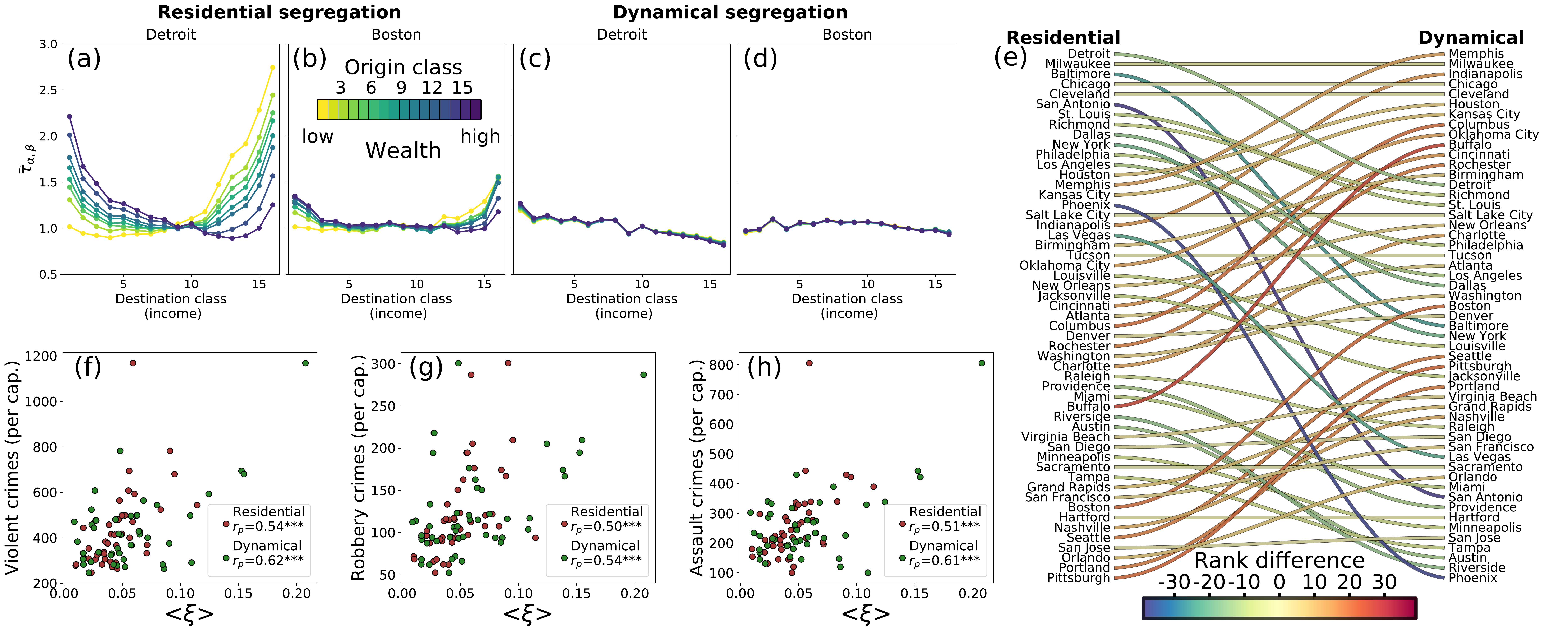} 
      \caption{Class mean first passage times and urban income
        inequality. (a)-(d) Class mean first passage times
        $\widetilde{\tau}_{\alpha \beta}$ between each of the income categories in
        Detroit and Boston when computed over the (a,b) adjacency
        (residential segregation) and the (c,d) commuting graphs
        (dynamical segregation. (e) Changes in the ranking of values
        between the residential and dynamical segregation. (f)-(h)
        Correlation between the index of segregation
        $\langle\xi\rangle$ and the criminality in US cities. (f)
        Violent crimes, (g) robbery crimes and (h) assault
        crimes.} \label{Figcities}
    \end{center}
\end{figure*} 

\section{Residential and dynamical urban economic segregation}

Socio-economic segregation has an enormous impact on city livability,
and many different measures to quantify it have been proposed in the
past years. However, most of those measures, with a few notable
exceptions~\cite{Barter2019}, focus on first-neighbour information,
and disregard the role of the mobility of
citizens~\cite{Winship1977,Reardon2004,Ballester2014,Louf2016}.  We
test here the potential of CMFPTs to quantify urban economic
segregation in large metropolitan areas, taking into account the daily
mobility patterns of individuals.

We considered the $53$ US cities with more than one million
inhabitants, and census information about the number of households in
each of the 16 income categories defined by the US Census Borough (see
Table~\ref{Table1} and the 2017 American Community
survey~\cite{income}), where class 1 is lowest income and class 16 is
the highest one. We constructed two different graphs among census
tracts, namely the graph of tract adjacency and the graph of daily
workplace commuting~\cite{Commuting}. The former graph is undirected
and unweighted, and is better suited to measure the so-called
residential segregation, i.e., the extent to which people with similar
levels of income tend to live in close-by areas. The commuting graph,
instead, is directed and weighted so that the weight of a link going
from $i$ to $j$ is given by the sum of the residents in $i$ working in
$j$ and the residents of $j$ working in $i$ in order to mimick the daily mobility of
citizens. 
  
The fact that households of more than one category are present in each
block group prevents us from assigning a single class to each unit, so
we computed the distribution of CMFPT by averaging over a large number
of realisations of class assignments. In each realisation, the class
of each node in the graph is sampled from the distribution of
household in the corresponding census tract, so that the probability
that node $i$ is assigned to class $\alpha$ in a given realisation is
equal to $m_{i, \alpha}/\sum_{\forall \beta} m_{i, \beta}$, where
$m_{i,\alpha}$ is the number of people of category $\alpha$ living in
node $i$. We took a slightly different approach for the commuting
graph as described in \cite{Bassolas2020b}, where the population 
attributed to node $i$ is a combination of the resident population 
at $i$ and the number of commuters working at $i$:
\begin{equation}
  \widetilde{m}_{i,\alpha}=m_{i,{\alpha}}+\sum_j
  \omega_{ji}\frac{m_{j,{\alpha}}}{\sum_{\forall \beta} m_{j, \beta}},
  \label{eq:norm_popul}
\end{equation}
where $\omega_{ji}$ is the weight from node $j$ to node $i$ in the
commuting graph, which is equal to the total number of people who live
in $j$ and commute to their workplace in $i$. By doing so we take into
account the fact that commuters effectively contribute to the
diversity of an area, as they spend a considerable amount of time in
there and actually interact with other commuters coming from different
areas as well. For each realisation of a class assignment, we run
$10^4-10^6$ random walkers from each of the nodes. Then, we obtain the
CMFPT $\tau_{\alpha \beta}$ for each ordered pairs of classes
($\alpha, \beta$) by averaging over all nodes and realisations of
class assignments, and we analyse the normalised CMFPT
$\widetilde{\tau}_{\alpha \beta}=\tau_{\alpha \beta}/\tau^{\rm
  null}_{\alpha \beta}$.

In Fig.~\ref{Figcities}(a)-(b) we show the profiles of
$\widetilde{\tau}_{\alpha \beta}$ from each of the 16 classes computed
over the adjacency graph, respectively for Detroit and Boston. It is
worth noting that the categories at the two extremes (i.e., the
poorest and the wealthiest ones) exhibit a quite similar pattern in
the two cities. They are both characterised by larger values of CMFPT
from any of the other classes, meaning that those two classes are in
general more isolated from the rest of the population, with most of
the high-income classes appearing slightly more isolated in Detroit
than in Boston. Moreover, in both cities all classes have a virtually
identical value of CMFPT to class 9, and very similar values to class
8 and 10, which indicates that these middle-income classes play a
pivotal role in the spatial distribution of income. However, despite
the fact that the qualitative behaviour is similar in the two cities,
there are some noticeable quantitative differences. First of all, the
values of normalised CMFPT are significantly larger in Detroit than in
Boston. Second, in Detroit we observe a strong dependence of CMFPT on
the class of the starting node, while in Boston the average number of
steps required to reach a class below 12 is almost constant regardless
of the category of the origin node. These important quantitative
differences would suggest that the spatial distribution of income in
Detroit is more heterogeneous than in Boston, a conclusion which is in
line with the classical literature about economic segregation in the
US \cite{Logan2013,pew,Waitzman,Jargowsky1996}.

However, in large metropolitan areas most of the daily activities of
individuals happen far away from their home, due to urbanisation
pressure and to decentralisation of productive sectors, so that
residential segregation can hardly tell the whole story. Indeed, in
the last years there has been an increasing interest in the
quantification of social and economic segregation by taking into
account mobility patterns
\cite{LeRoux2017,Petrovic2018,Randon2020}. This is easily doable
within our framework by letting the walkers move on the mobility graph
instead of the adjacency network between census tracts. We have
computed $\widetilde{\tau}_{\alpha \beta}$ upon the mobility graph of
each US city in the data set, as obtained from workplace commuting
information. The results are shown in Fig.~\ref{Figcities}(c)-(d), and
provide an interesting picture of the differences between residential
and mobility-focused segregation. First of all, in both cities
$\widetilde{\tau}_{\alpha\beta}$ does not depend too much on the
origin class $\alpha$ but instead on the destination class
$\beta$. This is most likely due to the fact that people of different
backgrounds commute to similar areas -- i.e. the city centre and
industrial sites--. Still, both cities display a distinct organisation
of CMFPTs. In Detroit low income classes are much more isolated
(higher $\widetilde{\tau}_{\alpha \beta}$) compared to high income
classes, which exhibit systematically lower values of
$\widetilde{\tau}_{\alpha \beta}$. In the case of Boston, instead,
$\widetilde{\tau}_{\alpha \beta}$ is almost flat with no important
dependence on either the source or the destination class.

The profiles of $\widetilde{\tau}_{\alpha \beta}$ that we show in
Fig.~\ref{Figcities}(a)-(d) provide an overall clear picture of the
distribution of CMFPT in a metropolitan area, but do not allow to
easily compare two cities in a systematic manner. Hence, we devised
two synthetic indices $\overline{\xi^{out}}$ and $\overline{\xi^{in}}$
that summarise the information on spatial income heterogeneity in a
single number. The idea behind these quantities is that an income
class $\alpha$ is more heterogeneously distributed if there is a large
difference between the CMFPT from $\alpha$ to the income classes
immediately adjacent to $\alpha$ and the median CMFPT from $\alpha$ to
any other class. We consider the discrepancy between the local and
global median of CMFPT from class $\alpha$:
\begin{equation*}
  \xi^{out}_\alpha = |\overline{\widetilde{\tau}}_{\alpha_{nn}}^{\rm out} -
  \overline{\widetilde{\tau}}^{\rm out}_{\alpha}|
\end{equation*}
where $\overline{\widetilde{\tau}}^{out}_\alpha$ is the median of
$\widetilde{\tau}_{\alpha \beta}$ when $\alpha\neq\beta$ and
$\overline{\widetilde{\tau}}_{\alpha_{nn}}^{\rm out}$ is the median of
$\widetilde{\tau}_{\alpha \beta}$ to its nearest neighbours
$\beta\in\{\alpha-1, \alpha, \alpha+1\}$~\footnote{If $\alpha=1$ or
  $\alpha=16$ we only consider the median between, respectively,
  $\{\alpha,\alpha+1\}$ or $\{\alpha-1,\alpha\}$.} Similarly for the
discrepancy between local and global median of CMFPT to class $\alpha$:
\begin{equation*}
  \xi^{in}_\alpha = |\overline{\widetilde{\tau}}_{\alpha_{nn}}^{\rm in} -
  \overline{\widetilde{\tau}}^{\rm in}_{\alpha}|
\end{equation*}
where $\overline{\widetilde{\tau}}^{\rm in}_\alpha$ and
$\overline{\widetilde{\tau}}_{\alpha_{nn}}^{\rm in}$ are now the
median of $\widetilde{\tau}_{\beta\alpha}$ when $\alpha\neq\beta$ and
the median of $\widetilde{\tau}_{\beta\alpha}$ from its nearest
neighbours $\beta\in\{\alpha-1, \alpha, \alpha+1\}$, respectively.
  
In Fig.~\ref{Figcities}(e) we show the ranking of US cities induced by
$\langle\xi\rangle=(\overline{\xi^{in}}+\overline{\xi^{out}})/2$, that
is the average discrepancy between local and global CMFPT from/to each
income class. In general, larger values of $\langle\xi\rangle$
indicate more pronounced levels of segregation. On the left-hand side
of the panel the cities are ranked according to $\langle\xi\rangle$ in
the adjacency graph of census tracts, while on the right-hand side
the ranking is based on $\langle\xi\rangle$ in the commuting
graph. Interestingly, Detroit is the first US city by residential
segregation, with other cities traditionally known for their high
levels of segregation like Milwaukee and Cleveland following
closely. Conversely, Boston is at the bottom of the ranking. However,
the ranking changes substantially if we consider instead the mobility
graph, and what we call dynamic segregation, as reported in the
right-hand side of Fig.~\ref{Figcities}(e). For instance, Baltimore
(which is ranked pretty high for residential segregation) gets
relegated to a mid-rank position, while cities like Buffalo or
Indianapolis, where residential segregation is not that high, get to
the top of the ranking of dynamic segregation.

The most interesting aspect of $\langle\xi\rangle$ is that it captures
some of the most undesirable consequences of income segregation,
namely the incidence of different types of crimes obtained from
\cite{ucr}. In Fig.~\ref{Figcities}(f)-(h) we report the correlation
between incidence per-capita of assaults, violent crimes, and
robberies with the levels of residential segregation measured on the
adjacency and on the commuting graph of the cities in our data
set. Interestingly, both indices display a significant correlation
with all three types of crime. Moreover, the indices computed over the
commuting network display a stronger correlation in all three cases,
reinforcing the idea that quantifying segregation by disregarding
mobility can indeed lead to distorted conclusions. For instance, the
city that appears on the top of the dynamical segregation ranking
(Memphis) is the one displaying the highest incidence of violent
crimes per-capita among the largest US cities, although it is placed
in the second quartile of the ranking by residential segregation. As a
comparison, we report in Appendix \ref{appendix:cities} the
correlations of traditional metrics of segregation (i.e., the Spatial
Gini coefficient and the Moran's I index) with the same crime
indicators, showing that they are not able to attain values as high as
those obtained with $\langle\xi\rangle$. Besides the lower
correlations, it is also important to note that both the Spatial Gini
Coefficient and the Moran's I index are symmetric quantities, and as
such they fail to capture the intrinsic asymmetry between income
classes revealed by Fig.~\ref{Figcities}(a)-(d).

Overall, our methodology based on the diffusion of random walks is not
only a natural extension of the latter multi-scalar approaches
introduced to characterise residential segregation~\cite{Olteanu2019},
but it also allows us to define a dynamical segregation that includes
mobility into the analysis as it has been recently discussed for
instance in Refs.~\cite{Le2017,Randon2020}.

\section{Conclusions}

Although being a relatively simple and intuitive tool, random walks
have been extensively and successfully used to model and characterise
transportation~\cite{De2014,Bassolas2020},
biological~\cite{Codling2008,Nicosia2017}, and financial
systems~\cite{Bacry2001}. They have proven useful to detect meaningful
structural properties in complex networks~\cite{Masuda2017,Zhang2013},
including communities~\cite{Pons2006,Rosvall2008,Lambiotte2014}, node
roles~\cite{Newman2005}, navigability~\cite{De2015}, and temporal
variability~\cite{Hoffmann2013}. Moreover, interesting insights about
the relation between the structure and dynamics on a network have come
from the analysis of transient and long trajectories of random walks
on graphs, including the statistics of first passage times and coverage
times~\cite{Redner2001,Condamin2005,Condamin2008,Fronczak2009,Bonaventura2014},
or the systematic study of their fluctuations~\cite{Nicosia2014}.
However, the potential usefulness of random walks to quantify the
heterogeneity of class distributions on networks, either in
spatially-embedded systems or in high-dimensional networks, has only
recently been hinted to~\cite{Nicosia2014,Bassolas2020b,Sousa2020}.

We have shown here that the information captured by the distribution
of inter-class mean first passage times can be used not just as a way
to detect the presence of anisotropy and correlations in the
properties of nodes, but also as a reliable proxy for the dynamics and
emergent behaviours of a complex system. One of the most interesting
aspects of the measures of heterogeneity, polarisation, and
segregation that we have introduced in this work is that they take
into account microscopic, meso-scopic and global relations among
classes, due to the fact that in principle random walks integrate
information about paths of all possible lengths. Another relevant
property of the measures of segregation based on CMFPT is that they
are non-parametric and correctly normalised with respect to a
meaningful null-model, hence allowing us to compare on equal grounds
the heterogeneity of class distributions in systems of different
sizes, which is where most of the classical indices of segregation
fail~\cite{Reardon2004,Olteanu2019}. Even more importantly, the
profiles of inter-class mean-first passage times are not symmetric
with respect to classes, and provide fine-grained information about
which classes are most responsible for the emergence of polarisation
and heterogeneity. In this respect, it would be worthy exploring how
the simple measure of polarisation that we proposed can be extended to
the case of more than two classes.

The fact that measures of class heterogeneity based on random walk
statistics correlate quite well with some of intrinsic dynamics
happening in social network (i.e., the spread of an epidemic) and with
some other exogenous processes mediated by the underlying graph (i.e.,
the incidence of crime in a city) confirm that the profiles of CMFPT
are a useful toolbox for targeted mitigation of the undesired effects
of these dynamics. For instance, the groups of a social network that
are more central according to $\widetilde{\tau}_{\alpha \beta}$ might
be the best candidates for early vaccinations to slow-down an
epidemic. At the same time, the definition of dynamical segregation
based on walks on the mobility graphs, and the fact that it correlates
quite substantially with crimes, potentially paves the way for a
re-definition of the traditional role attributed to residential
segregation, in favour of a more balanced view that takes into account
the activity patterns of citizens together with the spatial
distribution of their dwellings.

The generality of the methodology proposed in this paper, and its
applicability to different classical problems in complexity science,
establish a concrete link between classical statistical physics and
modern complexity science, and have the potential to provide new
interesting insights about the relation between structure and
dynamics of complex systems.

\begin{acknowledgments}
  The authors acknowledge support from the EPSRC New Investigator
  Award Grant No. EP/S027920/1. This work made use of the MidPLUS
  cluster, EPSRC Grant No. EP/K000128/1.
  \href{http://doi.org/10.5281/zenodo.438045}{doi.org/10.5281/zenodo.438045}.
\end{acknowledgments}

\appendix

\section{Derivation of CMFPT in simple geometries}

\subsection{Chain of nodes}
\label{appendix:chain}

We derive the full distribution of CMFPT for the chain of nodes
depicted in Fig.~\ref{fig:fig2}(b), as a function of the length $L$ of
the chain. For $L\ge 6$ the forward equations for the mean first
passage time to blank nodes read:
\begin{equation}
  T_{0,\bl} = 1 + \frac{1}{4} T_{1,\bl}
  \label{eq:chain_T0}
\end{equation}
\begin{equation}
  T_{M-k,\bl} = 1 + \frac{1}{4}\left[T_{M-k+1,\bl} + T_{M-k-1,\bl}\right], \quad
  k = 1,2,\ldots, M-1
  \label{eq:chain_TMk}
\end{equation}
\begin{equation}
  T_{M,\bl} = \left\{
    \begin{aligned}
      & \frac{4}{3} + \frac{1}{3}T_{M-1,bl} \quad\quad {\rm if}a \>L\>
      {\rm is\> even }\\
      & 1 + \frac{1}{2} T_{M-1,\bl} \quad\quad {\rm if}\>L\> {\rm is\> odd}
    \end{aligned}
    \right.
    \label{eq:chain_TM}
\end{equation}
By iteratively substituting $T_{M-k+1}$ into Eq.~(\ref{eq:chain_TMk})
for $k=1, 2, \ldots $ we find a parametric recursive expression for
$T_{M-k,0}$:
\begin{equation}
  T_{M-k,\bl} = A_k + B_k T_{M-k-1,\bl}
  \label{eq:chain_TMk_recur}
\end{equation}
Although the two sequences $A_k$ and $B_k$ depend on whether $L$ is
even or odd, as we explain below, it is possible to derive a generic
solution for this set of recurrence equations. In particular, we
write Eq.~(\ref{eq:chain_TMk_recur}) for $k=M-1$:
\begin{equation}
  T_{M-(M-1),\bl} = T_{1,\bl} = A_{M-1} + B_{M-1} T_{0,\bl}
  \label{eq:chain_T1}
\end{equation}
and we substitute it in Eq.~\ref{eq:chain_T0} to obtain:
\begin{align}
  T_{0,\bl} & = 1 + \frac{1}{4}T_{1,\bl} = 1 + \frac{1}{4}\left[A_{M-1} +
    B_{M-1} T_{0,\bl}\right]\nonumber\\
  & \Rightarrow T_{0,\bl} = \frac{4 + A_{M-1}}{4 - B_{M-1}}
  \label{eq:chain_T0_final}
\end{align}
By plugging Eq.~(\ref{eq:chain_T0_final}) into
Eq.~(\ref{eq:chain_T1}), and then recursively propagating back the
result in Eq.~(\ref{eq:chain_TMk_recur}), we obtain a closed
expression for $T_{k,\bl}$ when $k=0,1,\ldots M-1$:
\begin{equation}
  T_{k,\bl} =
  \left[\frac{4+A_{M-1}}{4-B_{M-1}}\right]\prod_{\ell=1}^{k}B_{M-\ell}
  + \sum_{j=1}^{k-1}A_{M-k+j}\prod_{\ell=0}^{j-1}B_{M-k+\ell}
\end{equation}
which together with Eq.~(\ref{eq:chain_TM}) provides the entire
distribution of $T_{k,\bl}$ across the chain. As we mentioned above,
the actual form of $A_k$ and $B_k$ depends on whether the number of
nodes $L$ in the chain is even or odd. If $L$ is \textit{even} we
have:
\begin{align*}
  A_{k} & = \frac{4a_k}{b_{k+1}}\\
  B_{k} & = \frac{b_{k}}{b_{k+1}}
\end{align*}
where
\begin{align*}
  a_{k} &= \{1,4,15,56,209,\ldots\}\\ b_{k} &=
  \{1,3,11,41,153,\ldots\}.
\end{align*}
It is easy to prove that both $a_k$ and $b_k$ satisfy the recurrent
relation:
\begin{equation}
  x_{n} = 4x_{n-1} - x_{n-2}, \quad n=0,1,\ldots
  \label{eq:generic_recur}
\end{equation}
with slightly different initial conditions. By solving these recurrent
relations we obtain the closed forms:
\begin{align}
  a_{k} &= \frac{\sqrt{3} + 2}{2\sqrt{3}}(2+\sqrt{3})^k +\frac{\sqrt{3}
    - 2}{2\sqrt{3}}(2-\sqrt{3})^k\\
  b_{k} &= \frac{\sqrt{3} + 1}{2\sqrt{3}}(2+\sqrt{3})^k +\frac{\sqrt{3}
    - 1}{2\sqrt{3}}(2-\sqrt{3})^k
\end{align}
which can be used to compute directly the values of $A_k$ and $B_k$;

In the case of \textit{odd} length $L$, instead, we obtain:
\begin{align*}
  A_{k} & = \frac{c_{k}e_{k}}{d_{k}f_{k}}\\ B_{k} & = \frac{e_{k}}{4f_{k}}
\end{align*}
where:
\begin{align*}
  c_k & = \{8,28,104,388,1448,\ldots\}\\
  d_k & = \{7, 26, 97, 362, 1351,\ldots\}\\
  e_k & = \{5,19,71,265,989,\ldots\}\\
  f_k & = \{4,14,52,194,724,\ldots\}
\end{align*}
Also these four sequences satisfy the general recurrence relation in
Eq.~(\ref{eq:generic_recur}), and by solving the relation for each of
the corresponding initial conditions we obtain the closed forms:
\begin{align}
  c_k &= \frac{8\sqrt{3} + 12}{2\sqrt{3}}(2 + \sqrt{3})^k +
  \frac{8\sqrt{3} - 12}{2\sqrt{3}}(2 - \sqrt{3})^k\\
  d_k &= \frac{7\sqrt{3} + 12}{2\sqrt{3}}(2 + \sqrt{3})^k +
  \frac{7\sqrt{3} - 12}{2\sqrt{3}}(2 - \sqrt{3})^k\\
  e_k &= \frac{5\sqrt{3} + 9}{2\sqrt{3}}(2 + \sqrt{3})^k +
  \frac{5\sqrt{3} - 9}{2\sqrt{3}}(2 - \sqrt{3})^k\\
  f_k &= \frac{4\sqrt{3} + 6}{2\sqrt{3}}(2 + \sqrt{3})^k +
  \frac{4\sqrt{3} - 6}{2\sqrt{3}}(2 - \sqrt{3})^k
\end{align}

The cases where $1\le L<6$ must be considered separately. If $L=1$,
then the chain degenerates in a single white node surrounded by four
black nodes. In this case, we simply have $T_{0,\bl} = 1$. If $L=2$,
instead, the chain contains only two nodes, and is equivalent to the
minimal weighted graph shown below:
\begin{center}
  \includegraphics[height=0.75in]{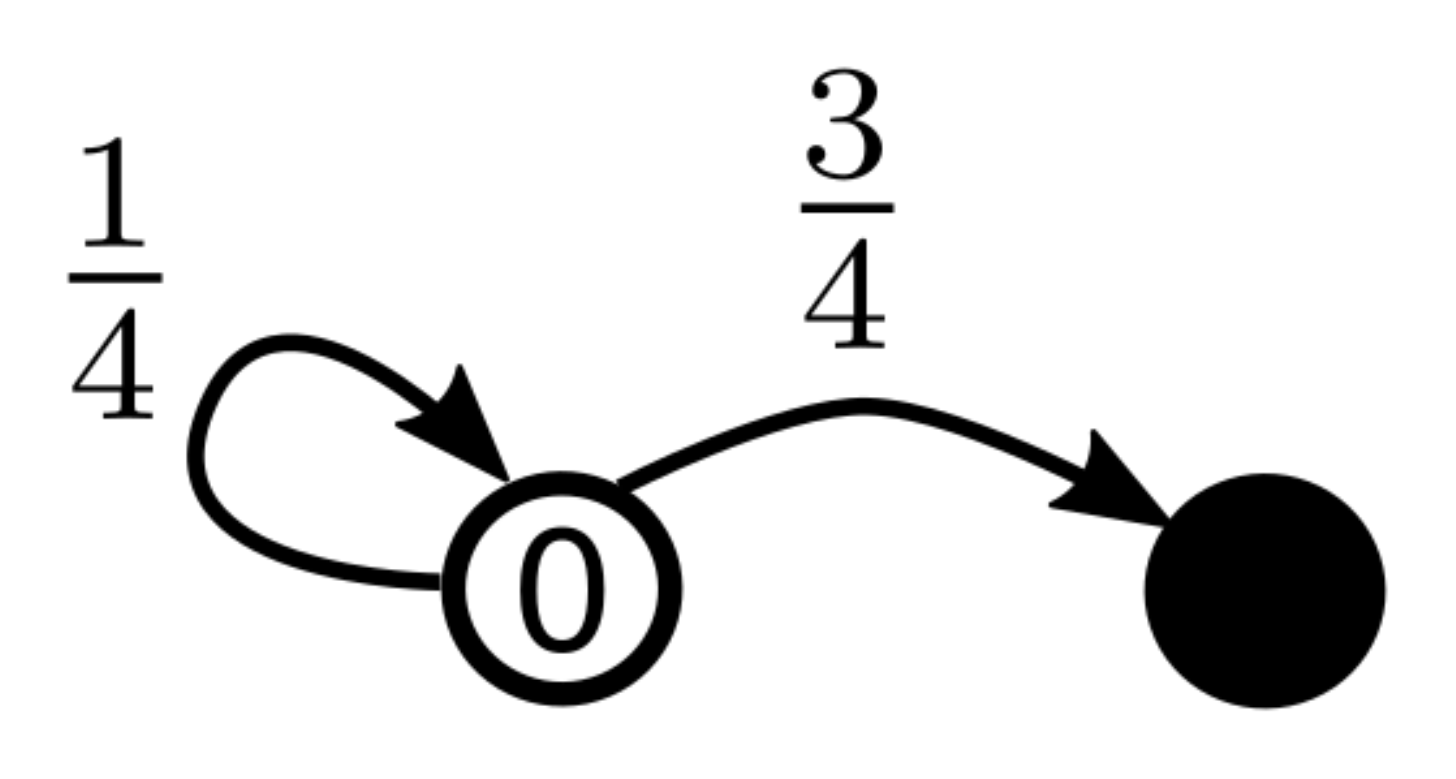}
\end{center}
where for convenience we have shown only the links coming out of white
nodes. The forward equation for node $0$ gives:
\begin{align*}
  T_{0,\bl} &= 1 + \frac{1}{4}T_{0,\bl}\\
  \Rightarrow & T_{0,\bl} = \frac{4}{3}
\end{align*}
The case $L=3$ is equivalent to the minimal graph:
\begin{center}
  \includegraphics[height=1.5in]{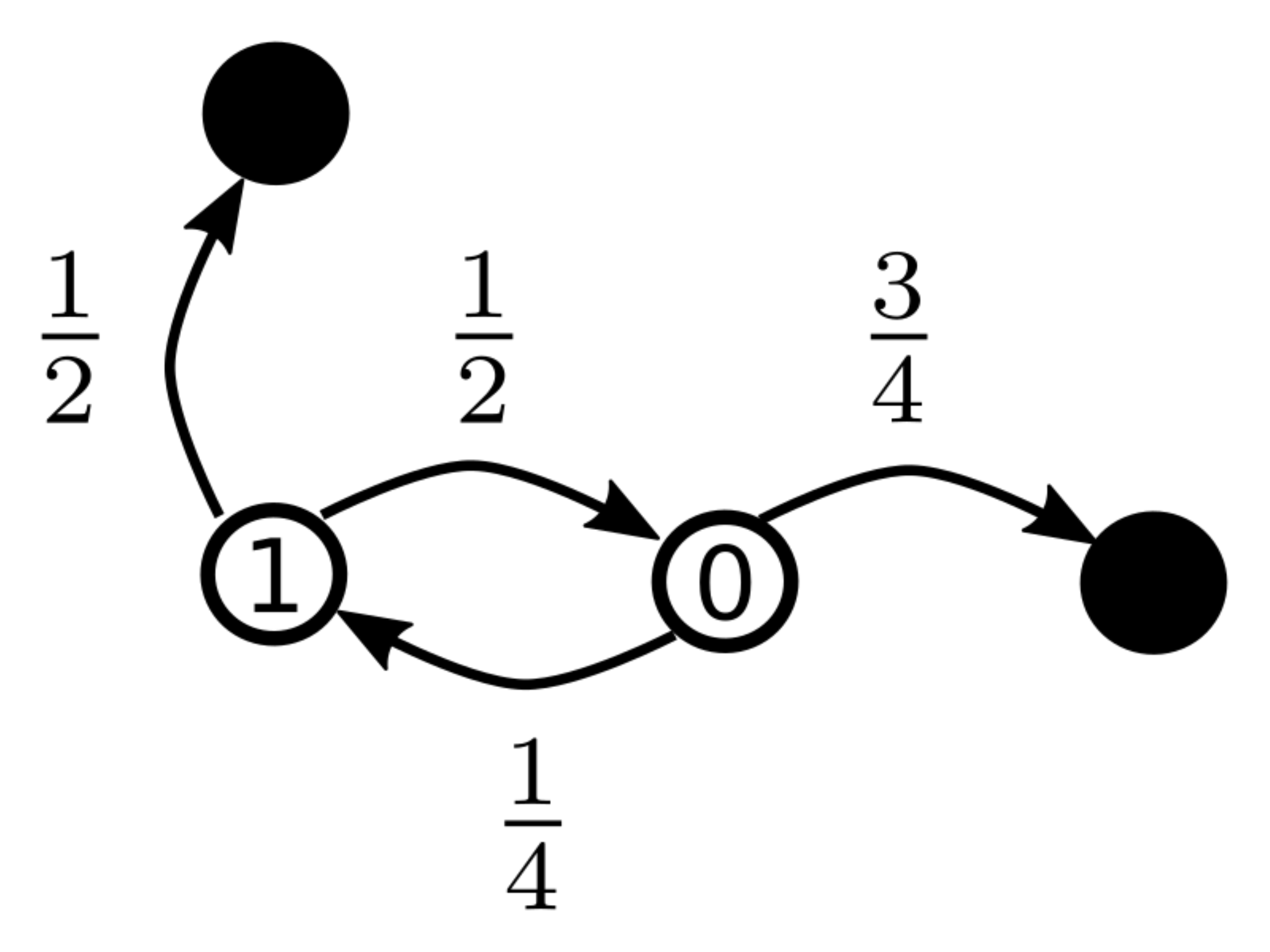}
\end{center}
and the two mean first passage times to black nodes satisfy the system
of equations:
\begin{equation}
  \left\{
  \begin{array}{rl}
    T_{0,\bl} & = 1 + \frac{1}{4}T_{1,\bl}\\
    T_{1,\bl} & = 1 + \frac{1}{2}T_{0,\bl}\\
  \end{array}
  \right.
\end{equation}
whose unique solution is:
\begin{equation}
  T_{0,\bl} = \frac{10}{7}, \quad\quad T_{1,\bl} = \frac{12}{7}
\end{equation}

When $L=4$, the minimal weighted graph reduces to:

\begin{center}
  \includegraphics[height=1.5in]{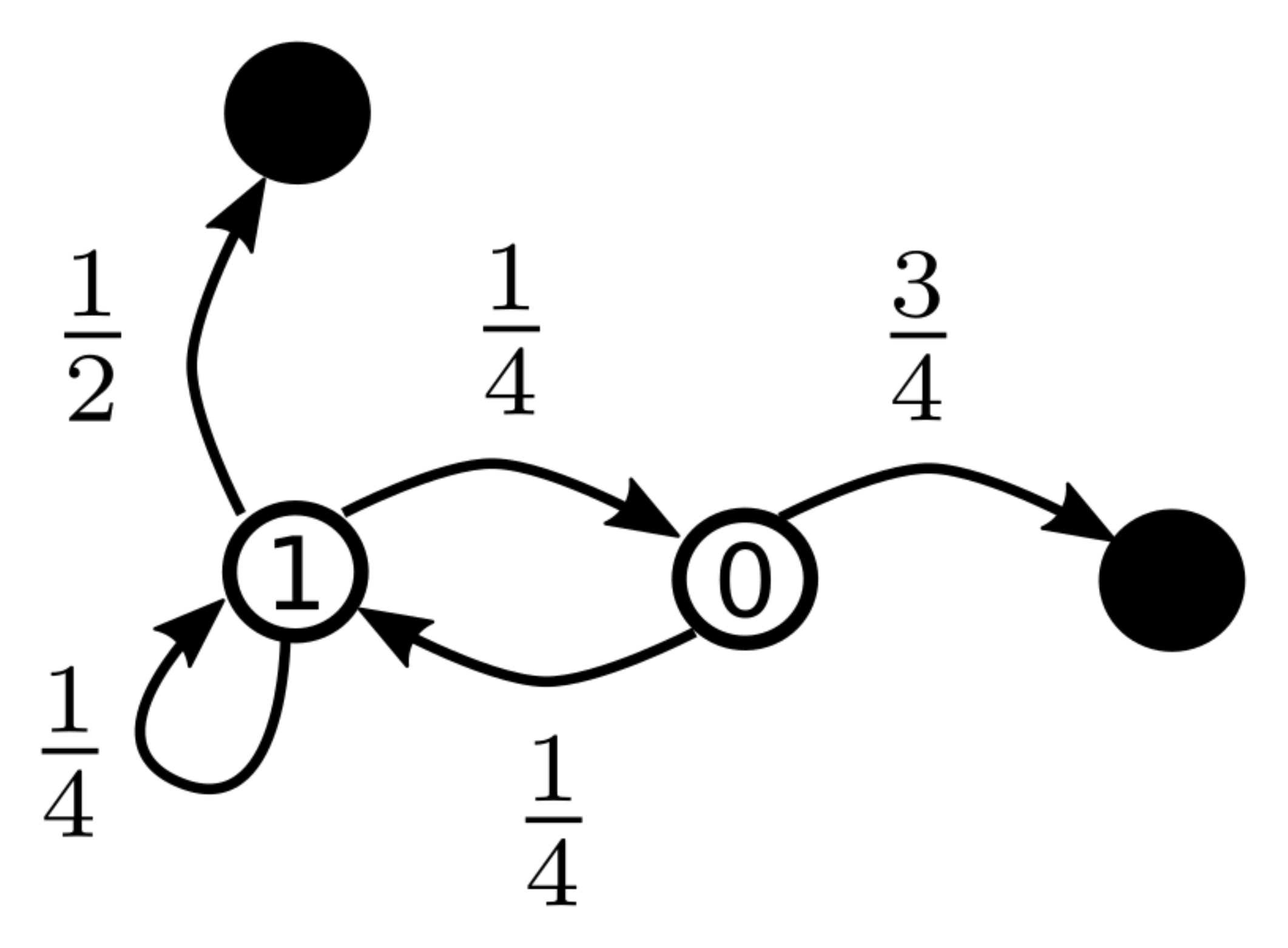}
\end{center}

and the two mean first passage times satisfy the equations:

\begin{equation}
  \left\{
  \begin{array}{rl}
    T_{0,\bl} & = 1 + \frac{1}{4}T_{1,\bl}\\
    T_{1,\bl} & = 1 + \frac{1}{4}\left(T_{0,\bl} + T_{1,\bl}\right)\\
  \end{array}
  \right.
\end{equation}
whose solution is:
\begin{equation}
    T_{0,\bl} = \frac{16}{11}, \quad\quad T_{1,\bl} = \frac{20}{11}
\end{equation}
Finally, for $L=5$ the minimal weighted graph is:

\begin{center}
  \includegraphics[height=1.5in]{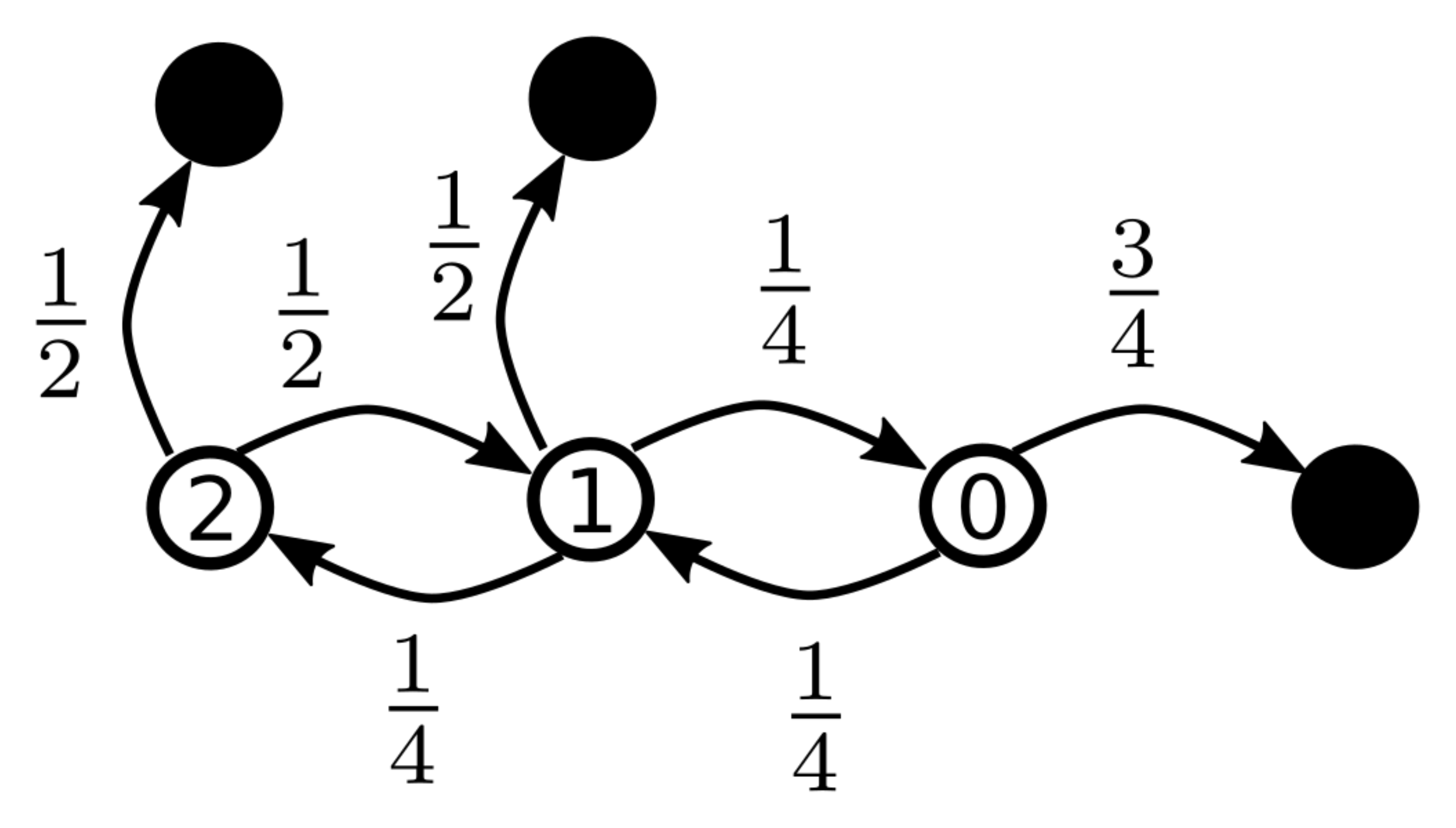}
\end{center}
which corresponds to the following system of equations:
\begin{equation}
  \left\{
  \begin{array}{rl}
    T_{0,\bl} & = 1 + \frac{1}{4}T_{1,\bl}\\
    T_{1,\bl} & = 1 + \frac{1}{4}\left(T_{0,\bl} + T_{2,\bl}\right)\\
    T_{2,\bl} & = 1 + \frac{1}{2}T_{1,\bl}
  \end{array}
  \right.
\end{equation}
and yields:
\begin{equation}
  T_{0,\bl} = \frac{19}{13}\quad\quad T_{1,\bl} =\frac{24}{13}
  \quad\quad T_{2,\bl} = \frac{25}{13}\quad\quad
\end{equation}

\subsection{Infinite cylinder}
\label{appendix:cylinder}

The distribution of CMFPT from white nodes to blank ones in the
infinite cylinder is derived by induction on the generic term
$T_{M-k,\bl}, \quad k=0,1,\ldots M$. In particular, when $k=0$ we
obtain the mean first-passage time from the farthest node $M$, whose
forward equation is:
\begin{equation}
  T_{M,\bl} = 1 + \frac{2}{3} T_{M,\bl} + \frac{1}{3}T_{M-1,\bl}
\end{equation}
which leads to:
\begin{equation}
  T_{M,\bl} = 3 + T_{M-1,\bl}
\end{equation}
which in turn depends only on $T_{M-1, \bl}$. If we now plug this
expression in the forward equation for $T_{M-1,\bl}$:
\begin{equation}
  T_{M-1, \bl} = 1 + \frac{1}{2}T_{M,\bl} + \frac{1}{2}T_{M-2,\bl}
\end{equation}
we obtain:
\begin{equation}
  T_{M-1, \bl} = 7 + T_{M-2, \bl},
\end{equation}
which in turn depends only on $T_{M-2,\bl}$. By considering the first
few terms of this sequence:
\begin{align*}
  T_{M-1,\bl} & = 7 + T_{M-2,\bl}\\
  T_{M-2,\bl} & = 11 + T_{M-3,\bl}\\
  T_{M-3,\bl} & = 15 + T_{M-4,\bl}\\
  T_{M-4,\bl} & = 19 + T_{M-5,\bl}\\
\end{align*}
we realise that the generic term $T_{M-k,\bl},\quad k=0,1,\ldots M-1$
satisfies the recurrence equation:
\begin{equation}
  T_{M-k,\bl} = (4k+3) + T_{M-k-1,\bl}.
  \label{eq:cyl_T_Mk}
\end{equation}
Notice that the forward equation for $T_{0,\bl}$ is somehow simpler,
and reads:
\begin{equation*}
  T_{0,\bl} = 1 + \frac{1}{2} T_{0,\bl} + \frac{1}{4} T_{1,\bl}
\end{equation*}
which can be rewritten as:
\begin{equation}
  T_{0,\bl} = 2 + \frac{1}{2}T_{1,\bl}.
  \label{eq:cyl_T_0}
\end{equation}
\begin{figure*}
    \begin{center}
    \includegraphics[width=18cm]{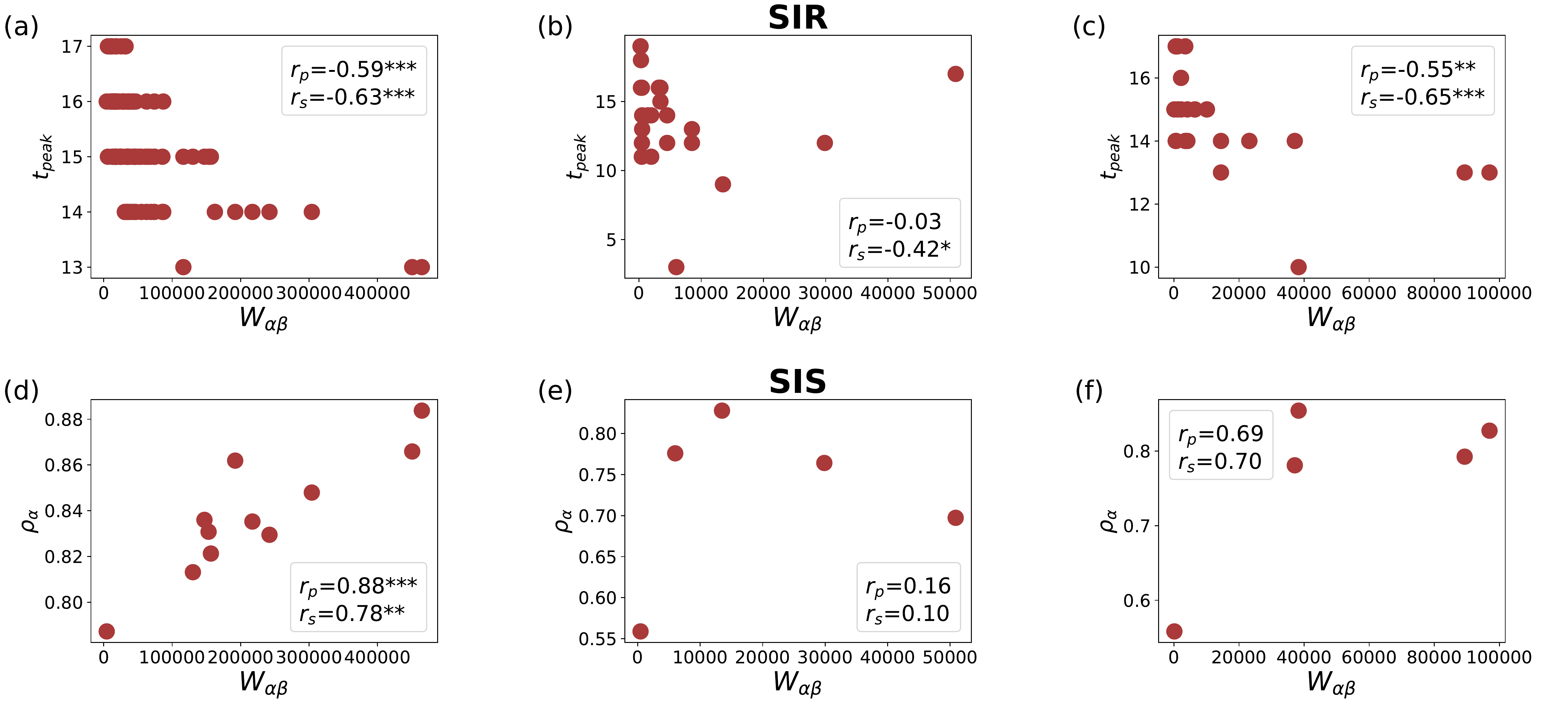}
      \caption{Epidemic spreading in contact networks and inter-group contacts.
(a)-(c) Connection between the total number of contacts between classes and the time steps until the peak of the epidemic in a SIR model.
(d)-(f) Connection between the total number of contacts within each class and the fraction of infected individuals in each class
in the stationary state.} \label{FigSIRtot}
    \end{center}
\end{figure*}
If we now consider Eq.~(\ref{eq:cyl_T_Mk}) for $k=M-1$:
\begin{equation}
  T_{1,\bl} = 4(M-1) + 3 + T_{0,\bl}
  \label{eq:cyl_T_1}
\end{equation}
and we plug it into Eq.~(\ref{eq:cyl_T_0}), we get:
\begin{equation}
  T_{0,\bl} = 2 + \frac{1}{2}\left[4(M-1) + 3 + T_{0,\bl}\right]
\end{equation}
which can be solved for $T_{0,\bl}$, yielding:
\begin{equation}
  T_{0,\bl} = 4M+3.
\end{equation}
By substituting $T_{0,\bl}$ back in Eq.~(\ref{eq:cyl_T_1}), and
recursively using the results to obtain $T_{2,\bl}, T_{3,\bl},
\ldots$, it is easy to find that:
\begin{equation}
  T_{k,\bl} = (k+1)(4M+3 -2k), \quad k=0,1,\ldots M-1
\end{equation}
and
\begin{equation}
  T_{M,\bl} = 3 + M(5 + 2M)
\end{equation}
$\Box$.

\begin{figure*}
    \begin{center}
    \includegraphics[width=18cm]{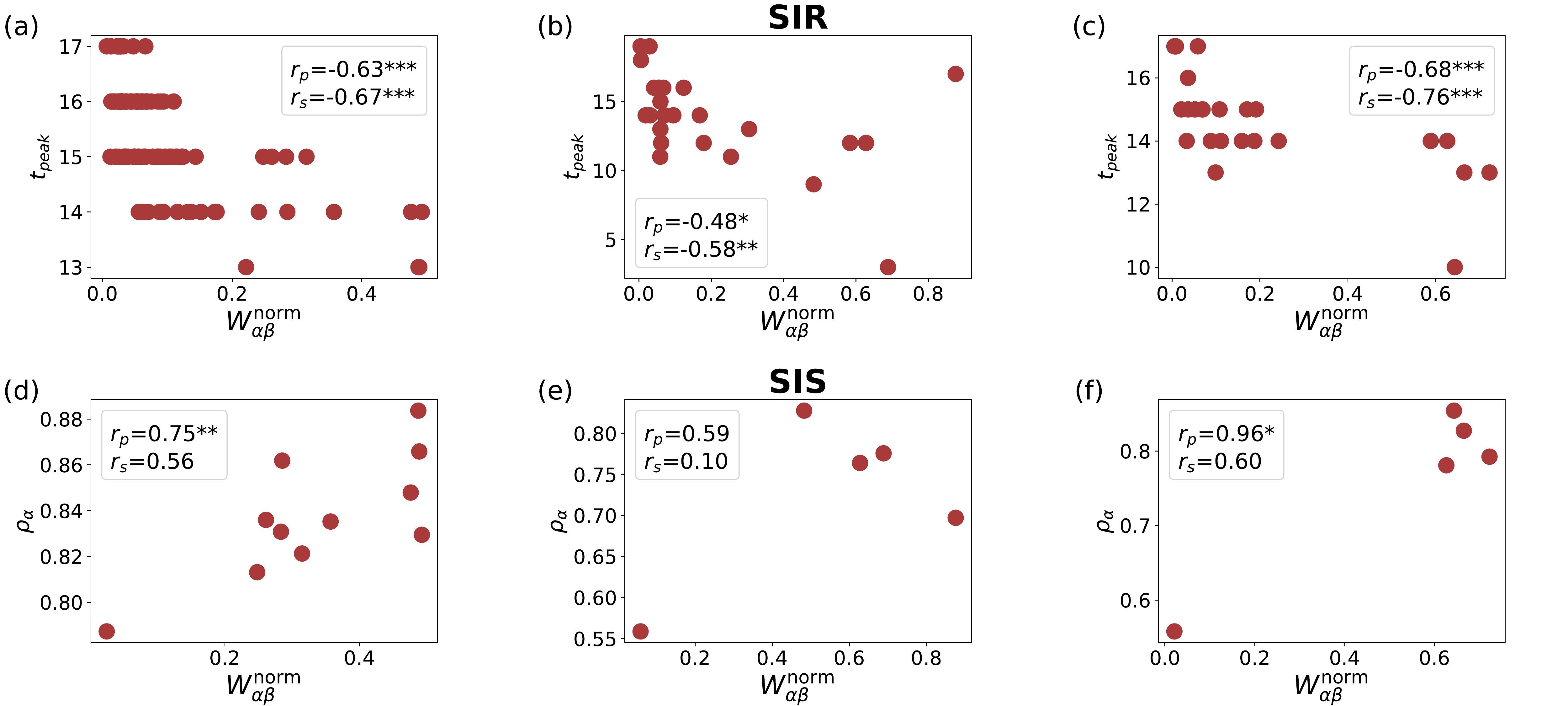}
      \caption{Epidemic spreading in contact networks and inter-group normalised contacts.
(a)-(c) Connection between the normalised contacts between classes and the time steps until the peak of the epidemic in a SIR model.
(d)-(f) Connection between the normalised contacts within each class and the fraction of infected individuals in each class
in the stationary state.} \label{FigSIRnorm}
    \end{center}
\end{figure*}

\section{Supplementary results for the epidemic spreading in contact networks}
\label{appendix:epidemics}

We report here the correlation between some simple structural measures
and the same epidemic variables considered in Fig.~\ref{FigSIR}. In
particular, we have considered two main quantities, namely, the total
number of contacts between individuals of group $\alpha$ and group
$\beta$ and the normalised contacts of the members of group $\alpha$
with those of group $\beta$. The first quantity is symmetric and is
just the sum of the total number of contacts between the members:
\begin{equation}
W_{\alpha \beta}=\sum_{ij}w_{ij}\delta_{f_i,\alpha}\delta_{f_j,\beta},
\end{equation}
while the second one:
\begin{equation}
W^{\rm norm}_{\alpha \beta}=\frac{W_{\alpha\beta}}{\sum_{\beta}W_{\alpha\beta}},
\end{equation}
is normalised by the total number of contacts of the group of origin
and, therefore, is in general not symmetric.

In Fig.~\ref{FigSIRtot} we show the correlation between the total
number of steps until the peak and the total number of contacts
between each pair of groups, as well as the correlation between the
fraction of population in each group infected in the endemic state
(SIS model) and the total contacts between members of the same group.
Despite correlations are relatively high and significant in some
cases, none of them is higher than those obtained for
$\widetilde{\tau}_{\alpha \beta}$.

Similarly, we show in Fig.~\ref{FigSIRnorm} the correlation of both
the SIR and the SIS models with the normalised number of
contacts. Correlations look slightly higher than in the case of the
total number of contacts, yet they are still considerablu lower than
the values obtained for
$\widetilde{\tau}_{\alpha\alpha}$. Particularly in the case of the SIS
model (stationary dynamics), we do not observe the clear alignment
that appears in Fig.~\ref{FigSIR}.

In Fig.~\ref{FigSIR} we reported the correlation coefficients for only
$3$ of the $7$ contact networks we have analysed. Table~\ref{Table2}
includes additional results of Pearson's $r_p$ and Spearman's $r_s$
for many combinations of the parameters and a for wider variety of
contact networks. Interestingly, the results are compatible with those
reported in Fig.~\ref{FigSIR}, and confirm that correlations between
CMFPT and critical epidemiological variable are large and
significative.

\begin{table*}[ht!]
\begin{tabular}{|l|l|l|l|l|l|l|l|l|}
\hline
 \hline
$\beta$ & $\mu$ & Network & Model & Pearson's $r_p$ & Spearman's $r_s$ & Model & Pearson's $r_p$ & Spearman's $r_s$  \\
  \hline
0.5 & 0.1 & Thiers13 & SIR & 0.92*** & 0.85*** & SIS & -0.30 & -0.67* \\  \hline
0.5 & 0.1 & InVS15 & SIR & 0.47*** & 0.73*** & SIS & -0.86*** & -0.89*** \\  \hline
0.5 & 0.1 & LH10 & SIR & 0.70*** & 0.86*** & SIS & -0.67 & -0.90* \\  \hline
0.5 & 0.1 & InVS13 & SIR & 0.29 & 0.73*** & SIS & -0.99** & -1.00*** \\  \hline
0.5 & 0.1 & Hospital & SIR & 0.89*** & 0.94*** & SIS & -0.89 & -1.00*** \\  \hline
0.5 & 0.1 & Enterprise & SIR & 0.59** & 0.71*** & SIS & -0.95* & -0.70 \\  \hline
0.5 & 0.1 & LyonSchool & SIR & 0.69*** & 0.73*** & SIS & -0.96*** & -0.89*** \\  \hline
0.5 & 0.3 & Thiers13 & SIR & 0.88*** & 0.79*** & SIS & -0.62 & -0.73* \\  \hline
0.5 & 0.3 & InVS15 & SIR & 0.33*** & 0.37*** & SIS & -0.27 & -0.52 \\  \hline
0.5 & 0.3 & LH10 & SIR & 0.57** & 0.59** & SIS & -0.95* & -0.90* \\  \hline
0.5 & 0.3 & InVS13 & SIR & 0.68*** & 0.70*** & SIS & -0.54 & -0.90* \\  \hline
0.5 & 0.3 & Hospital & SIR & 0.30 & 0.50* & SIS & -0.97* & -1.00*** \\  \hline
0.5 & 0.3 & Enterprise & SIR & 0.62*** & 0.68*** & SIS & 0.48 & 0.00 \\  \hline
0.5 & 0.3 & LyonSchool & SIR & 0.50*** & 0.55*** & SIS & -0.96*** & -0.95*** \\  \hline
0.9 & 0.1 & Thiers13 & SIR & 0.93*** & 0.86*** & SIS & -0.20 & -0.55 \\  \hline
0.9 & 0.1 & InVS15 & SIR & 0.47*** & 0.74*** & SIS & -0.82** & -0.80** \\  \hline
0.9 & 0.1 & LH10 & SIR & 0.68*** & 0.83*** & SIS & -0.66 & -0.90* \\  \hline
0.9 & 0.1 & InVS13 & SIR & 0.66*** & 0.68*** & SIS & -0.98** & -1.00*** \\  \hline
0.9 & 0.1 & Hospital & SIR & 0.77*** & 0.80*** & SIS & -0.87 & -0.80 \\  \hline
0.9 & 0.1 & Enterprise & SIR & 0.51** & 0.56** & SIS & -0.95* & -0.70 \\  \hline
0.9 & 0.1 & LyonSchool & SIR & 0.67*** & 0.73*** & SIS & -0.94*** & -0.86*** \\  \hline
0.7 & 0.3 & Thiers13 & SIR & 0.93*** & 0.87*** & SIS & -0.47 & -0.63 \\  \hline
0.7 & 0.3 & InVS15 & SIR & 0.28*** & 0.47*** & SIS & -0.67* & -0.59* \\  \hline
0.7 & 0.3 & LH10 & SIR & 0.65*** & 0.76*** & SIS & -0.84 & -0.90* \\  \hline
0.7 & 0.3 & InVS13 & SIR & 0.52** & 0.68*** & SIS & -0.98** & -1.00*** \\  \hline
0.7 & 0.3 & Hospital & SIR & 0.63** & 0.83*** & SIS & -0.95 & -1.00*** \\  \hline
0.7 & 0.3 & Enterprise & SIR & 0.67*** & 0.69*** & SIS & -0.93* & -0.90* \\  \hline
0.7 & 0.3 & LyonSchool & SIR & 0.34*** & 0.55*** & SIS & -0.97*** & -0.93*** \\  \hline
0.7 & 0.1 & Thiers13 & SIR & 0.93*** & 0.86*** & SIS & -0.25 & -0.67* \\  \hline
0.7 & 0.1 & InVS15 & SIR & 0.45*** & 0.72*** & SIS & -0.84*** & -0.85*** \\  \hline
0.7 & 0.1 & LH10 & SIR & 0.62*** & 0.77*** & SIS & -0.65 & -0.90* \\  \hline
0.7 & 0.1 & InVS13 & SIR & 0.75*** & 0.79*** & SIS & -0.99** & -1.00*** \\  \hline
0.7 & 0.1 & Hospital & SIR & 0.86*** & 0.89*** & SIS & -0.87 & -0.80 \\  \hline
0.7 & 0.1 & Enterprise & SIR & 0.61** & 0.64*** & SIS & -0.95* & -0.70 \\  \hline
0.7 & 0.1 & LyonSchool & SIR & 0.64*** & 0.69*** & SIS & -0.95*** & -0.86*** \\  \hline
0.9 & 0.5 & Thiers13 & SIR & 0.89*** & 0.78*** & SIS & -0.56 & -0.63 \\  \hline
0.9 & 0.5 & InVS15 & SIR & 0.31*** & 0.40*** & SIS & -0.38 & -0.50 \\  \hline
0.9 & 0.5 & LH10 & SIR & 0.58** & 0.76*** & SIS & -0.92* & -0.80 \\  \hline
0.9 & 0.5 & InVS13 & SIR & 0.62*** & 0.69*** & SIS & -0.89* & -1.00*** \\  \hline
0.9 & 0.5 & Hospital & SIR & 0.29 & 0.58* & SIS & -0.97* & -1.00*** \\  \hline
0.9 & 0.5 & Enterprise & SIR & 0.68*** & 0.72*** & SIS & -0.35 & -0.70 \\  \hline
0.9 & 0.3 & Thiers13 & SIR & 0.92*** & 0.85*** & SIS & -0.39 & -0.58 \\  \hline
0.9 & 0.3 & InVS15 & SIR & 0.36*** & 0.64*** & SIS & -0.81** & -0.81** \\  \hline
0.9 & 0.3 & LH10 & SIR & 0.61** & 0.77*** & SIS & -0.68 & -0.90* \\  \hline
0.9 & 0.3 & InVS13 & SIR & 0.50* & 0.71*** & SIS & -0.99** & -1.00*** \\  \hline
0.9 & 0.3 & Hospital & SIR & 0.59* & 0.94*** & SIS & -0.92 & -1.00*** \\  \hline
0.9 & 0.3 & Enterprise & SIR & 0.65*** & 0.73*** & SIS & -0.96* & -0.70 \\  \hline
0.7 & 0.5 & Thiers13 & SIR & 0.85*** & 0.73*** & SIS & -0.73* & -0.78* \\  \hline
0.7 & 0.5 & InVS15 & SIR & 0.35*** & 0.35*** & SIS & 0.01 & -0.41 \\  \hline
0.7 & 0.5 & LH10 & SIR & 0.61** & 0.47* & SIS & -0.87 & -0.80 \\  \hline
0.7 & 0.5 & InVS13 & SIR & 0.77*** & 0.61** & SIS & 0.64 & 0.40 \\  \hline
0.7 & 0.5 & Hospital & SIR & 0.50 & 0.55* & SIS & -0.89 & -0.80 \\  \hline
0.7 & 0.5 & Enterprise & SIR & 0.64*** & 0.73*** & SIS & 0.72 & 0.10 \\  \hline
0.7 & 0.5 & LyonSchool & SIR & 0.53*** & 0.53*** & SIS & -0.91*** & -0.98*** \\  \hline
\end{tabular}
\caption[Inc\\ ]{Pearson correlation coefficient $r_p$ and Spearman's
  rank correlation coefficient $r_s$ for the seven contact networks
  analysed for both the SIR and SIS models.} \label{Table2}
\end{table*}

\section{Supplementary results in the analysis of urban inequalities}
  \label{appendix:cities}

We report in Table~\ref{Table1} the income intervals that correspond
to each of the classes investigated in the case of urban inequalities.

\begin{table}[ht!]
\begin{tabular}{llll}
\hline
 \hline
 Class & Income & Class & Income  \\
  \hline
Class 1 & Less than $10,000$ &  Class  9 &   $45,000$ to $49,999$  \\ \hline 
Class 2 & $10,000$ to $14,999$ &  Class 10  &    $50,000$ to $59,999$  \\ \hline 
Class 3 &  $15,000$ to $19,999$ &  Class 11 &      $60,000$ to $74,999$  \\ \hline 
Class 4 &  $20,000$ to $24,999$  &   Class 12 &      $75,000$ to $99,999$  \\ \hline 
Class 5 &  $25,000$ to $29,999$ &  Class 13 &    $100,000$ to $124,999$   \\ \hline 
Class 6 & $30,000$ to $34,999$  &   Class 14 &   $125,000$ to $149,999$   \\ \hline 
Class 7 &  $35,000$ to $39,999$  &  Class  15 &   $150,000$ to $199,999$    \\ \hline 
Class 8 &  $40,000$ to $44,999$  &  Class  16 &    $200,000$ or more   \\ \hline 
\end{tabular}
\caption[Income classes]{Corresponding income branch of each class or category in US dollars.} \label{Table1}
\end{table}

\subsection{Correlations with traditional segregation indicators}

\begin{figure}
    \begin{center}
    \includegraphics[width=8cm]{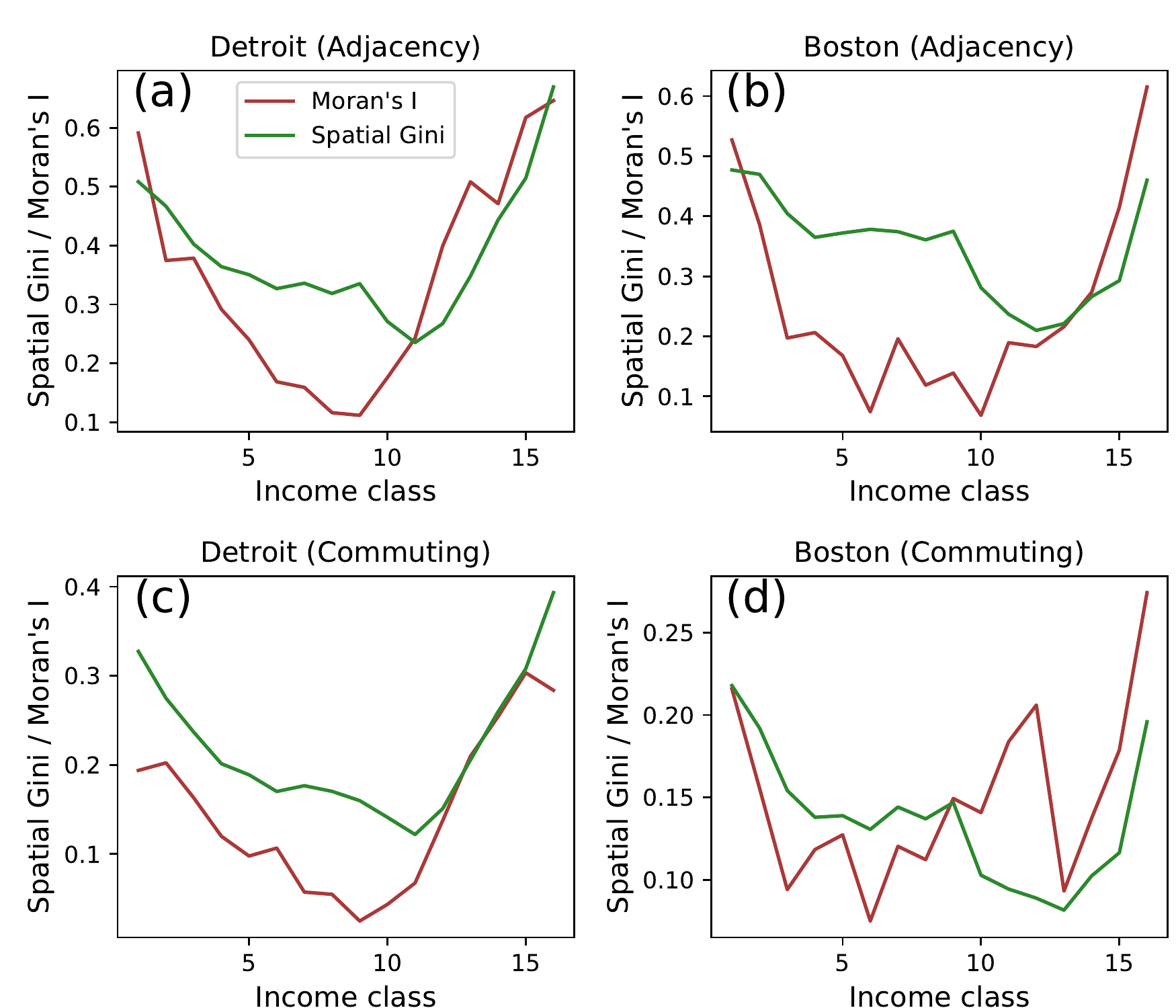}
      \caption{Segregation of each income category with traditional segregation indices when calculated
  on (a)-(b) the adjacency graph and (c)-(d) the commuting graph} \label{Segcities}
    \end{center}
\end{figure}

Given the correlations between the index constructed from CMFPT and
set of crime indicators in American cities, we have also checked if
traditional segregation indicators provide a comparable amount of
information.  The two main quantities we analysed are the Spatial Gini
and the Moran's I auto-correlation. We used of The Spatial Gini which
can be written as
\begin{equation}
SG=\frac{\sum^N_{i=1}\sum^N_{j=1}(1-w_{ij})|x_i-x_j|}{2N^2<x>},
\end{equation}
This metric is designed for binary networks as it is the case for the
adjacency graph of census tracts, however it is not defined for
non-binary weights as it is the case of the commuting graph. To take
into account those weights, we have modified it as
\begin{equation}
SG_w=\frac{\sum^N_{i=1}\sum^N_{j=1}(1-w_{ij}/w_{\rm max})|x_i-x_j|}{2N^2<x>},
\end{equation}
where $w_{\rm max}$ corresponds to the maximum weight observed in the network.
In the case of the Moran's I, it is given by
\begin{equation}
I=\frac{N}{\sum_{i}\sum_{j}w_{ij}}\frac{\sum_{i}\sum_{j}w_{ij}(x_i-<x>)(x_j-<x>)}{\sum_{i}(x_i-<x>)^2},
\end{equation}
which is generally defined for any type of network. However. there is
a slight difference between the adjacency and the commuting graph. In
the case of the adjacency graph we have used a row normalised approach
for the weights so that each cell counts the same in the global
average, while for the commuting graph no normalisation was applied so
that a cell with more (in)outgoing links will count more. In
Fig.~\ref{Segcities}, we provide the values of both indices for each
of the income categories analysed throughout this paper. As can been
seen, the behaviour observed is quite comparable to the one in
Fig.~\ref{Figcities}, with the low and high-income classes displaying
significantly higher segregation. Moreover, segregation seems to
decrease when computed over the commuting graph.

\begin{figure}
    \begin{center}
    \includegraphics[width=8.5cm]{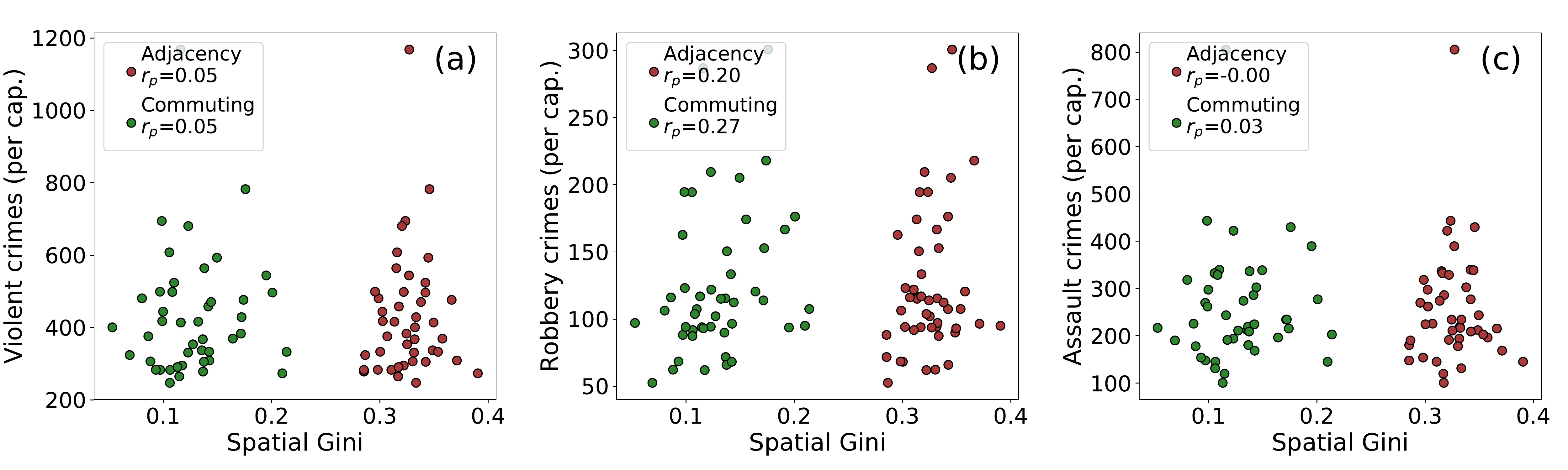}
      \caption{Economic segregation and criminality.
 (a)-(c) Correlation between the Spatial Gini and the criminality in US cities. (a)
Violent crimes, (b) robbery crimes and (c) assault crimes.} \label{Corrgini}
    \end{center}
\end{figure}

\begin{figure}
    \begin{center}
    \includegraphics[width=8.5cm]{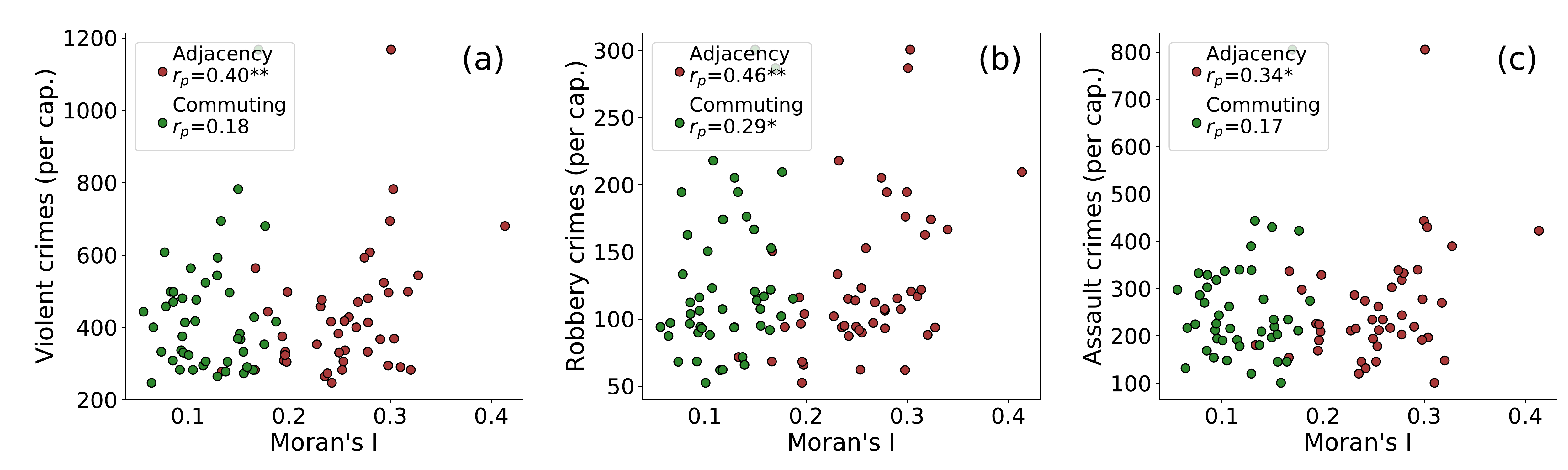}
      \caption{Economic segregation and criminality.
 (a)-(c) Correlation between the Moran's I index and the criminality in US cities. (a)
Violent crimes, (b) robbery crimes and (c) assault crimes.} \label{Corrmoran}
    \end{center}
\end{figure}

To extract a single metric for each of the cities studied, we have
taken the median over all income categories and we have compared the
correlations with those observed for the index extracted from
CMFPT. We report in Fig.~\ref{Corrgini} and Fig.\ref{Corrmoran} the
correlations for both quantities, the Spatial Gini and the Moran's I,
respectively. While a significant correlation appears in some cases,
the highest one observed for the Moran's I calculated on the adjacency
graph is still lower than the lowest correlations observed with our
index $\langle\xi\rangle$. This confirms the higher explicative power
of a metric based on the diffusion on random walks and the crucial
role of including mobility into the analysis.

\end{document}